\documentclass[twocolumns]{aa}
\usepackage{graphicx}
\begin{document}
\titlerunning{TSMS in the MCs. I. FUSE, HST, and VLT observations.}
\authorrunning{M.K. Andr\'e et al.}

   \title{Tiny--Scale Molecular Structures in The Magellanic Clouds.}
   \subtitle{I. FUSE, HST and VLT Observations.} 
   \author{M.K. Andr\'e \inst{1}, F. Le~Petit \inst{2,4}, P. Sonnentrucker \inst{3},
          R. Ferlet \inst{1}, E. Roueff \inst{2}, T. Civeit \inst{1}, J-.M. D\'esert \inst{1}, S. Lacour \inst{1},
	  A. Vidal--Madjar\inst{1}}

   \offprints{M.K. Andr\'e}

   \institute{Institut d'Astrophysique de Paris, CNRS/UPMC,
              98, bvd ARAGO, 75014 Paris\\
              \email{andre@iap.fr}
        \and
                LUTH and FRE2462 du CNRS, Observatoire de Paris, Place J. Jansen, 921995 Meudon Cedex, France
        \and            
                Johns Hopkins University
	\and
	        Onsala Space Observatory - Se 439 92 - Sweden          
   }

   \abstract{
  The objective of this series of two papers is to investigate small-scale molecular
  structures in the Magellanic Clouds (hereafter MCs).
We report on the {\small FUSE} detections of the HD and CO molecules {\bf on the lines of sight
towards three Large Magellanic stars}: Sk\,$-$67D05, Sk\,$-$68D135, and Sk\,$-$69D246. HD is also detected for the
  first time {\bf on the lines of sight towards two Small
  Magellanic Cloud stars}: AV\,95 and Sk\,159. While the HD and CO abundances are expected
  to be lower in the Large Magellanic Cloud where molecular fractions are a third of
  the Galactic value and where the photodissociation flux is up to thousands times larger,
  we report an average HD/H$_2$ ratio of 1.4$\pm$0.5 ppm and CO/H$_2$ ratio ranging from 0.8 to 2.7 ppm
  similar to the Galactic ones. We tentatively identify a
  deuterium reservoir (hereafter D--reservoir) towards the Small Magellanic Cloud, along the light path to AV\,95. We derive a D/H
  ratio ranging from 1. 10$^{-6}$ to 1.1 10$^{-5}$.
Combining {\small FUSE} and {\small HST}/{\small STIS} data we also analyzed 
  the H$_2$, Cl\,{\small I}, Cl\,{\small II}, Fe\,{\small II}, S\,{\small II}, C\,{\small I}, C\,{\small I}*, and
  C\,{\small I}** content, when available.
  High resolution {\small VLT} observations of Na\,{\small I}, Ca\,{\small II}, and Ca\,{\small I} were obtained in support of
  the lower resolution {\small FUSE} and {\small STIS} data for three targets in order to unravel the sightline velocity structures.
  These observations form the only such set of
  detections in the Magellanic Clouds to date and allow us to investigate in {\bf detail} some of the
  physical properties of the intervening molecular gas.
  Our observation of the HD and CO molecules in the Magellanic Clouds is an argument for
  dense ({\bf $n_{\rm H}$} $>$ 100 cm$^{-3}$) components. Furthermore, we demonstrate that these components
  are probably extremely small molecular clumps (possibly as small as 10$^2$ pc) or filaments similar
  to the tiny--scale atomic structures (TSAS) recently observed in the halo of our Galaxy by {\bf Richter et al. (2003)}. 
Interestingly enough, for these five sightlines, we also detected molecular hydrogen originating
  in the Galactic disk. From these observations, we conclude that tiny--scale molecular
  filamentary structures are present in the disk of the Galaxy as well.

   }

   \maketitle
%

\section{Introduction}

        For many years, the study of molecular clouds through the CO emission lines has greatly improved our knowledge of the
physical properties within the Magellanic Clouds themselves and helped constrain the theoretical
models of molecular clouds formation and chemistry in different Galactic environments
(Cohen et al. 1988). Over the past decade, in particular, the
Swedish--ESO--Submillimeter--Telescope (SEST)
has been used to map the radio--emission of molecular complexes within the LMC at an $8.\arcsec0$  angular resolution which is an order of magnitude 
higher compared to previous surveys. Such resolution corresponds to a linear extent of about 10 pc at the LMC distance of 50 kpc.
The latter survey led to the clear detection of small CO clumps with sizes possibly smaller than 10
pc (Lequeux et al. 1994; Garay et al. 2002), typical densities of a few thousand molecules per cubic centimeter and
virial masses around 10$^{4}$ solar masses. However, smaller clumps and AU-scale
structures, such as those tentatively identified in absorption in the Milky Way (Frail et al. 1994;
Lauroesch, Meyer \& Blades 2000; Richter, Sembach \& Howk 2003)
are still out of reach.

        An alternative method relies on detailed spectroscopic analyses of selected sightlines 
that might actually help to unravel such small-scale structures in the Magellanic
InterStellar Medium (MC ISM). The high-resolution spectroscopic abundance studies of
Welty et al. (1999) based on the observations of SN 1987A by Vidal-Madjar et al. (1987) showed that 
the MC ISM exhibits properties similar to those of the warm, low-density Galactic disk clouds and can contain complex velocity
structures with more than 20 components. Some of the absorbers observed
toward SN 1987A also showed signatures of relatively cool gas {\bf ($T<1500$ K)}.  The lack of information concerning
the molecular content precluded to further characterize this cool phase, however.

        With the {\small FUSE}  data it is now possible to investigate the molecular content of
this cool phase as well via H$_2$, CO, and HD.
Although the HD molecule is the most abundant molecule in the interstellar medium after H$_2$, relatively few observations have been
reported to date in the Galaxy and few comparisons with theoretical models of translucent clouds have been done (Le Petit et al. 2002).
Similarly to H$_2$, the {\bf presence of} HD molecule is best determined through
its ground state transitions which only occur in the FUV domain. The detection is however difficult due to a low molecular abundance
(HD/H$_2$ $\approx$ 10$^{-6}$). The ground-state transitions of HD were first observed with {\it Copernicus} in
the local ISM towards bright, nearby stars (Spitzer \& Cochran 1973; Wright \& Morton
1979). Only recently did the {\small FUSE} mission (Moos et al. 2000) re--open
the FUV window with sufficiently high sensitivity to allow the detection of HD further
away in the Galactic disk (Ferlet et al. 2000; Rachford et al. 2002; Lacour et al. 2004 in prep.). As to extragalactic search for HD, only
two detections were found in the literature, one in the LMC by Bluhm \& de Boer (2001) with
the {\small FUSE} satellite, the other in a Damped
Lyman Alpha system towards PKS1232$+$0.82 (Varshalovich et al. 2001). The conclusions from the work of Bluhm \& de Boer (2001)
are consistent with the existence of a cool phase in the LMC with molecular abundances (H$_2$, HD, CO) typical of Galactic
translucent sightlines (Rachford et al. 2002).
        
        The project of this series of two papers is to provide a more extensive knowledge
of the ``dense'', small-scale molecular structures within the Magellanic Clouds. We wish to point
out that we specifically use the term ``dense'' to designate gas with proton density, $n_{\rm H}$, above 100 cm$^{-3}$ in this work.
In the present paper, emphasis is put
on the analysis of H$_2$, HD, CO, Cl\,{\small I}, Cl\,{\small II}, Na\,{\small I},
Ca\,{\small II}, S\,{\small II}, Fe\,{\small II}, C\,{\small I}, C\,{\small I*}, and C\,{\small I**}, when available, 
towards five bright Magellanic Cloud stars.  Sk\,$-$67D05,
Sk\,$-$68D135 and Sk\,$-$69D246 are located in the LMC. AV\,95, and Sk\,159 are in the SMC. In $\S$ 2 we describe the reduction of the {\small FUSE}, {\small VLT} and {\small STIS} datasets. The analysis methods used to derive the species column densities are described in $\S$ 3.
Individual conclusions for each target are presented in $\S$ 4. A preliminary discussion of
these results is given in $\S$ 5 and $\S$ 6 summarizes our study.

\section{Observations}
\subsection{{\small FUSE} Observations}

With the help of the web--based archive maintained at the Institut d'Astrophysique de {\bf Paris}\footnote{More at http://fuse.iap.fr} we have searched
for HD lines among the 55 LMC sightlines and 26 SMC sightlines observed with {\small FUSE} (see the spectral atlas by
Danforth et al. 2002).
Each dataset was reprocessed with the CALFUSE pipeline version 2.0.5.
The reduction and calibration procedure includes
thermal drift correction, geometric distortion correction,
heliocentric velocity correction, dead time correction, wavelength
and flux calibrations (Sahnow et al. 2000).

The full observation of a given  star is often divided into
several sub-exposures which are usually aligned by cross-correlation over the entire
wavelength domain. However, due to the time--dependent stretching of the wavelength calibration
between exposure, coaddition of the entire spectra at once may lead to systematic shifts in
the final coadded spectrum. In other words, even though the reference absorption features are perfectly aligned
at one end of the spectrum, the global coaddition could result in random shifts at the other end of it. In
order to avoid these effects and optimize the coaddition process with the {\small FUSE} data, we divided each subexposure into 10 \AA~ windows and performed the coaddition within these spectral domains, weighting each domain by its exposure time.  No coaddition of data from different channels was performed. We however used each channel, independently, in the profile fitting procedure.
As a byproduct we computed a basic two components velocity structure (MW+MC) and overplotted a model of the molecular species on top of each spectrum
in order to detect the weak HD lines at the MCs velocities.

Finally, out of the 81 MC sightlines observed only five showed detectable amounts of HD; three detections occured in the LMC (Fig. 1) and two detections occured in the SMC (Fig. 2). It is important
to note that HD is possibly present in a few others lines of sight but could not
be detected because of insufficient S/N ratios or large amount of Galactic
molecular hydrogen crowding the HD lines arising from the MCs. In Table 1,
 we report the main stellar and interstellar parameters for each of the five stars. A summary of the {\small FUSE} observations is given in Table 2.

\subsection{{\small VLT}  Observations}

Spectra of the three LMC targets were obtained at the Very Large Telescope Kueyen at Paranal,
Chile, in 2003 April using the Ultraviolet Echelle Spectrograph ({\small UVES}, Dekker et al. 2000) in
visitor mode (see Table 3). The red and blue arm of {\small UVES}, respectively centered at 3900 \AA\ and 8600 \AA, were used at the same time in dichroic mode during the same night. A non-standard setup was used with the red arm centered at 7500 \AA.  We used the $0\farcs4$ slit to achieve a resolving power of 110,000 (2.6 pixels). With these setups we observed the Na\,{\small I} (5890 \AA) doublet, Ca\,{\small I} (4230 \AA), and Ca\,{\small II} (3930 \AA) doublet.
1-D spectra were extracted with the {\small UVES} pipeline (Ballester et al. 2000). A careful examination of the corrected 2D images showed that the
background and the potential stray light were well corrected with the standard pipeline procedure. However, we noted a shift of the order
of 14 km s$^{-1}$ in the wavelength calibration that was corrected using the telluric Na\,{\small I} emission lines (located at 0 km s$^{-1}$
geocentric velocity). S/N ratios of about 30 per pixel were achieved in about 20 minutes.

In order to remove the telluric lines that contaminate the
spectra around the Na\,{\small I} doublet (5890 \AA), we took an exposure of the well known early type, unreddened star $\alpha$ Vir.
We normalized the continuum of all the spectra to unity. We finally divided the normalized LMC spectra by the normalized $\alpha$ Vir spectrum. The
absence of telluric lines in the final spectra shown in Figs. 3, 4 and 5 are indicative of the effectiveness and accuracy of our method.

\subsection{{\small STIS} Observations}

One LMC star (Sk\,$-$69D246) and one SMC star (AV\,95) were previously observed with the {\small HST}/{\small STIS} instrument using the 140M grating and MAMA detectors (Woodgate et al. 1998; Kimble et al. 1998). The observations contain a complete wavelength coverage from 1150 \AA\ to 1730 \AA\ at 45,800 resolving power (2.0 pixels). The S/N ratios per pixel are about 15 and 10 for Sk\,$-$69D246 and
AV\,95, respectively.
We used these medium-resolution (6.5 km s$^{-1}$) echelle observations to derive
the column densities of C\,{\small I}, C\,{\small I}*, C\,{\small I}**, Cl\,{\small I}, and S\,{\small II} towards both sightlines. A summary of the {\small STIS} observations is given in Table 4.

\section{Line-of-sight velocity structure}

With the exception of the pathlength towards Sk\,159 for which little information is available, the detailed interstellar
component velocity structure is inferred from high resolution Na\,{\small I}, Ca\,{\small II}, or
Cl\,{\small I} spectra and is given in the heliocentric reference frame. The use of different type of tracers
allowed us to better differentiate between cold molecular or purely diffuse atomic components.

\subsection{Diffuse gas tracers: Ca\,{\small II}, Ca\,{\small I}, and Na\,{\small I}}

We observed the three LMC targets with the {\small VLT} with a resolving power of 110,000
around 589 nm, 423 nm, and 393 nm. These observations gave us access to three important
species that are usually used in combination to derive the physical conditions of
a sightline in the ISM.

The Na\,{\small I}/Ca\,{\small II} ratio for instance can be used as a tracer of the
integrated depletion (Welty et al. 1999). Typical values for this ratio
in the warm diffuse medium are of the order of $-$2.0 dex where the Ca is not depleted.
In colder, denser clouds (ignoring the shocks), two processes compete to drastically change the Ca\,{\small II}
abundance: a severe depletion of calcium onto the grains (by about 2.0 dex) and the
decrease of the ionization radiation field that makes of Ca\,{\small II} the dominant
ionization stage (Pottasch 1972; Bertin et al. 1993; Welty et al. 1996). In general,
in such cold media the Na\,{\small I}/Ca\,{\small II} is found to be as high as 2.5 dex
and more. We use this ratio to discriminate between possible diffuse atomic
or denser molecular components in the {\small VLT} data, thus better constraining the location
of the molecular absorbers seen with {\small FUSE} at lower resolution.

Ca\,{\small I} is clearly detected towards one of the LMC target: Sk\,$-$69D246.
In the cooler neutral gas phase where Ca\,{\small II} is the dominant ion stage,
the Ca\,{\small I}/Na\,{\small I} ratio can be a useful indicator of the temperature.
For $T$ $<$ 3000 K, $\log$ $N$(Ca\,{\small I})/$N$(Na\,{\small I}) is of the order of $-$0.1.
At temperatures lower than 1500 K, $\log$ $N$(Ca\,{\small I})/$N$(Na\,{\small I}) is in between
$-$2.5 and $-$4.5 (Welty et al. 2001). Moreover, if we assume that the ionization equilibrium is
reached in the plasma we can, in principle, use the ratio $N$(Ca\,{\small I})/$N$(Ca\,{\small II})
to get a direct estimate of $n_e$. In the case of
the calcium atoms, the relation is simply given by (Welty et al. 1999): $N$(Ca\,{\small I})/$N$(Ca\,{\small II}) $=$ 
$n_e$/$(\Gamma/\alpha)$ where $\Gamma$ is the photoionization rate and $\alpha$ is the
recombination coefficient.

\subsection{Molecular gas tracers: Cl\,{\small I}}

Chlorine possesses a unique chemistry in which Cl\,{\small II} reacts rapidely with  H$_2$ when the latter is optically thick to form HCl$^+$, which in turn leads to Cl\,{\small I} and H\,{\small I} (Jura 1974; Jura \& York
1978). Therefore, chlorine is predominantly ionized in H I regions while it is predominantly neutral when cold, optically thick H$_2$ components are present (Sonnentrucker et al. 2002, 2003). 

So far, only four Cl\,{\small I} lines have experimentally determined $f$-values (Schectman et al. 1993). Two of these lines (1347 and 1363 \AA) appear in the {\small STIS} range and the two others (1088 and 1097 \AA) in the {\small FUSE} range. 
We have positive detections of Cl\,{\small I} at 1088 \AA\ for all three LMC stars. Because the 1097 \AA\ line in the LMC gas is often blended with the Galactic components our Cl\,{\small I} column density estimate relies on the 1088 \AA\ line alone for  Sk\,$-$67D05, Sk\,$-$68D135 and Sk\,159. The 1347 \AA\ line was additionally used to constrain the Cl\,{\small I} column densities towards the two other stars for which archival {\small STIS} data exist. 

Cl\,{\small II} exhibits only two transitions, both in the {\small FUSE} range at 1063 \AA\ and 1071 \AA. The 1063 \AA\ line is often completely obliterated by molecular hydrogen, and hence unusable. The 1071 \AA, on the other hand, shows only mild-to-moderate blending with an adjacent H$_2$ ($J$ = 4) line and can therefore be used depending on the sightline complexity. 
However, the continuum around the latter line is lowered by a broad wind feature. This results in a reduced S/N ratio in that particular spectral region of the spectrum for both the LMC and SMC stars. We, therefore, only report one
secure detection of this line towards Sk\,$-$69D246 and an upper limit towards Sk\,$-$67D05.

\section{Analysis}

\subsection{Profile fitting (PF)}

Because of the velocity overlap between the Galactic and the MC features and the low resolution of the {\small FUSE} data, many absorption lines are blended. 
Extensive use of profile fitting methods is therefore mandatory to account for these blends. 

The profile fitting program, called {\it Owens}, is a fortran procedure developed by the {\small FUSE} french PI team
and is one of the most suitable for studying FUV data (Lemoine et al. 2002). The great advantage of this routine is the ability to fit
different spectral domains and various species, simultaneously. Thus, a good $b$-value estimate can be gained from the comparison of the Doppler broadenings
of heavy and light species while, at the same time, blends are naturally resolved. In practice, however, the atomic and
molecular species are rarely distributed in a similar way over the same components and a great caution must be used when
grouping species together in a single component.

Profile fitting of the atomic and molecular species were thus performed in two independant steps since the atomic absorbers usually show
much more complicated velocity structures and larger $b$-values (more details are given in subsequent sections).
Also, data from the different instruments were used at the same time in order to
fit simultaneously different species and constrain the different contributions to the doppler
broadening (temperature and turbulent velocity). Table 5 is a summary of selected atomic and molecular transitions
used in this work.

The other advantage of the {\it Owens} routine is that it allows to treat as parameters the instrument background, the Point Spread Function (PSF) and
possible shifts between spectral windows due to residual errors in the wavelength calibration.

In particular, special care must be taken as to the choice of the PSF when working with the
{\small FUSE} data. Despite many attempts to characterize the {\small FUSE} PSF (H\'ebrard et al. 2002, Wood et al.
2002), its true shape seems elusive and varying with time. Current values found in the {\small FUSE} literature
range from a quite optimistic LWRS FWHM of 15 km s$^{-1}$ to 25 km s$^{-1}$. In our case,
we have to coadd invidual exposures of the same channel and finally the composite PSF relies
on the quality of the coaddition process as well. However, because the S/N of the final data is relatively
low, we found that a simple gaussian PSF with 20 km s$^{-1}$ (resolution of 15,000) FWHM worked
well and various tests with FWHM 10 \% larger or 10 \% smaller showed that the PSF plays a minor role in the
final error budget largely dominated by the errors on the $b$-value. Throughout this paper we thus assume the
PSF to be independent of the dataset and the observation date.

The errors were derived with the standard $\chi^2$ method as described in H\'ebrard et al. (2002)
and
as shown in Fig. 6. We want to outline that, in most of the fits performed, the
converged $\overline{\chi^2}$
are close to unity indicating both that the fit are consistent with the data and that the data
error vector is
well estimated by the {\small FUSE} pipeline version 2.0.5. However, when the errors are dominated by
systematics, a more conservative method needs to be employed. Such was the case for the LMC H$_2$
$J$ = 2, 3 levels that are dominated by small errors in the $b$-value. For these, we explored the
different $b$-values consistent with the global fit and report the full range of possible
values.

\subsection{Apparent Optical Depth (AOD)}

Both AOD and profile fitting techniques were used to determine $N$(HD) and $N$(Cl\,{\small I}) in order
to better understand potential sources of systematic error. The AOD method (Savage \& Sembach
1991) yields apparent column densities, $N_a$, that are equivalent to the
true column density if no unresolved saturated structure is present. If such structure is present,
then $N_{\rm a}$ $<$ $N_{\rm PF}$, and the apparent column density is a lower limit to the true column density.
Our AOD analysis, including estimation of the errors, follows the procedures outlined in Savage \&
Sembach (1991). Note that it was not possible to apply the AOD technique to CO because of a severe
blend with one H$_2$ $J=$ 4 Galactic line. This blend was naturally resolved using the profile fitting technique.

\subsection{H$_2$ column densities}

We adopted the wavelengths, oscillator strengths, and damping constants
for the molecular hydrogen transitions from Abgrall et al. (1993a) for
the Lyman system and Abgrall et al. (1993b) for the Werner system. 
The inverses of the total radiative lifetimes are reported in Abgrall et al. (2000).
The profile fitting was performed both in the MW and the MCs over 10 to
20 different spectral regions simultaneously. Rotational levels up
to $J=$ 6 were considered.

In the MCs, the $J=$ 0 and $J=$ 1 levels are the most populated levels and show damping wings. The estimates of
the total H$_2$ column densities in the MCs are therefore well constrained by these two levels.
The higher $J$ levels, on the other hand, appear to show various degrees of saturation. Thus, prior knowledge of
the sightlines velocity
structure is mandatory to the analysis of these levels. This preliminary analysis
was performed using higher resolution (3 km s$^{-1}$) {\small VLT} optical data of Na\,{\small I}, Ca\,{\small II} that trace
the most diffuse components and {\small STIS} high resolution data of Cl\,{\small I} that traces the optically thick H$_2$ components.
Indeed, we find that with the exception of Sk\,$-$69D246 and Sk\,159, the present sightlines are dominated by a single molecular
absorber.

The errors were derived following the procedure described in $\S$ 3.1. Note however that while cold unresolved components
might well still be present in the {\small FUSE} spectra, our conclusions based on $J=$ 0 and $J=$ 1 lines remain
unaffected. The total H$_2$ column density and the kinetic temperature of this gas ($T_{01}$) still remain accurately
determined.

All of the molecular absorbers detected in the Galaxy
have column densities several orders of magnitude smaller than towards the Magellanic
Clouds. That is obviously an observational bias due to the severe blending between
the Galactic and the Magellanic features. An upper limit to the Galactic H$_2$ is
of the order of $\log$ $N$(H$_2$) $\approx$ 18.0 dex. Beyond that limit, Magellanic absorption
lines start to suffer {\bf from} the strong MW blending.

\subsection{HD column density}

All the ground state ro-vibrational bands of the HD molecules lie below 1180 \AA. The {\small FUSE}
range is therefore perfectly suited to study this molecule. 
The wavelengths for the $J=$ 0 and 1 rotational levels were taken from Roueff \& Zeippen
(1999) while the $f$-values come from the calculations
of Abgrall \& Roueff (2003). This new set of improved $f$-values takes into
account the rotational coupling between electronic states.

In the LMC, because of the severe blending of most of the HD lines
(shifted by about $+$250 km s$^{-1}$) with strong H$_2$ features from the Milky Way, we could only
select
a handful of useful spectral domains containing HD ($J=$ 0): around 959 \AA,
1011 \AA, 1043 \AA\ and 1066 \AA. Yet, depending on the S/N ratio
of a given spectrum and depending on the particular blending between
Galactic species and HD along a given sightline, some of these domains had
to be discarded for the analysis. As an example, profile fitting of HD in the spectrum of 
Sk\,$-$69D246 is shown in Fig. 5.

Our method of investigation combines AOD measurements
of the strongest detected transition (HD $J=$ 0 at 1043 \AA) with profile fitting performed
simultaneously over several spectral domains.
Both results are indeed comparable and demonstrate that the HD lines are on the linear part of the curve of growth. Thus,
the HD column is not sensitive to the $b$-value used in the profile fitting method.
Each of these methods allows us to get independent measurements, with
their own error estimates, that are then compared and averaged.

The situation is the same with HD lines originating in the SMC with the
difference that the typical velocity shift is only $+$150 km s$^{-1}$.
Towards the SMC, there are only three available HD lines clear of blending at 1007 \AA, 1011 \AA,
and 1043 \AA. Again, the final
result is the average of a profile fitting and an AOD
measurements.

For the sightline towards Sk\,$-$69D246, we compared our result with the work of Bluhm \& de Boer (2001) towards the LMC. Our
HD column density  appears to be only half of their
claimed value. While they used only 2 transitions of the HD molecule
(the 1043 \AA\ and 1066 \AA\ lines) assuming no blending, we fitted 2 additional
transitions at 1011 \AA\ and 959 \AA\ with {\small FUSE}. Performing simultaneous fits over several windows, we were able to unravel a blend of the
1066 \AA\ line. Indeed, the equivalent width measured at 1066 \AA~ is 
twice as big as expected from the measurement of the equivalent width
at 1043 \AA. To our knowledge, this extra absorption feature does not correspond to
any known species either at the MW or at the LMC velocity. Finally we attributed this blend
to a stellar line (see Fig. 7). This blend
was, hence, responsible for an overestimation of the HD column density
published by Bluhm \& de Boer (2001).

\subsection{CO}

Two strong $^{12}$CO bands are usually detected in the {\small FUSE} spectra, the E-X band at 1076 \AA\ and the $^{12}$CO C-X band
centered at 1087 \AA. We detected the 1087 \AA\ band in the spectra of the LMC sightlines while we tentatively
derived upper limits towards the SMC sightlines. The  other bands present in the spectra could not be used either
because they are too weak or because of blends with a stronger Galactic feature. The $f$-values were taken from
Morton \& Noreau (1994) for the 1087 \AA\ line.

The $^{13}$CO absorption lines are also located in the {\small FUSE} range and are
totally blended with the $^{12}$CO
absorption lines. However, we did not add the $^{13}$CO bands in our profile fitting
since the $^{13}$C isotopes are supposed to be a minute fraction
of the total carbon atoms. It is well known that the $^{12}$C/$^{13}$C ratio is
a sensitive indicator of the degree of stellar nucleosynthesis and is expected to
decrease with the Galactic chemical evolution (Woosley \& Weaver 1995). Since
the $^{12}$CO/$^{13}$CO ratio is of the order of 70 in the Galaxy (Savage et al. 2002) it is clear that $^{13}$CO is negligible at the level of accuracy of our analysis.

Our fits include rotational levels up to $J=$ 3. Despite the weakness of the band we conducted an investigation of the individual population distribution. Noteworthy is that the CO
rotational levels spread over 100 km s$^{-1}$ in the C-X band.
It is, hence, possible to partially deblend their different participation to the
profile with the {\small FUSE} resolution ($\approx$ 20 km s$^{-1}$). Rotational
levels greater than $J=$ 4 are found to be negligible which is
consistent with subthermal excitation temperatures ($\approx$ 4 K) measured in the diffuse Galactic disk (Rachford et al. 2002).

Note that an additional difficulty towards the LMC stars resides in the partial blending
of the 1087 \AA\ line with a Galactic H$_2$ $J=$ 4 line and a Cl\,{\small I} line. Fortunately the analysis of the Galactic
H$_2$ and Cl\,{\small I} makes it possible to safely deblend the LMC CO absorption lines when present.

\subsection{C\,{\small I}, C\,{\small I}*, and C\,{\small I}**}

C\,{\small I} absorption lines are present in all our {\small FUSE} spectra. However, given that most of these
lines are multiplets and that the S/N ratio is not high enough to deblend the absorption lines, it is not possible
to derive the relative population of the excited states. The situation is worsened by the potential blend
between Galactic and MC C\,{\small I} absorption lines.

 In the case of Sk\,$-$69D246 and AV\,95, however, the blending can
be resolved due to the existence of complementary {\small STIS} data that also contain a wealth of carbon lines at higher resolution
(see Table 5). We estimate the column densities for the carbon lines and use them to derive the average density
of the probed clouds.

The relative population of neutral carbon atoms in excited fine-structure levels of the ground electronic state can be
used to estimate the number densities and temperatures of neutral clouds in the ISM (Jenkins \& Shaya 1979; Jenkins \&
Tripp 2001). Using the $f_1$--$f_2$ diagram ($f_1$ $=$ $N$(C\,{\small I}*)/$N$(C\,{\small I}$_{\rm total}$);
$f_2$ $= N$(C\,{\small I}**)/$N$(C\,{\small I}$_{\rm total}$)), we thus derive the thermal pressures ($P/k$) towards AV\,95 and Sk-69D246.
Assuming that the carbon lines are arising in the dense molecular component with  a temperature $T$ $=$ $T_{01}$(H$_2$) we can then infer the average H number density, $n_{\rm H}$ of the medium crossed (see Table 6 and sections below).

\section{Results for individual sightlines}

\subsection{Sk\,$-$67D05}

Sk\,$-$67D05 is a O9.7Ib star (Neubig, Bruhweiler \& Smith 1999; Tumlinson et al. 2002) located near the
Western edge of the LMC. It lies in a diffuse H II region (Chu et al. 1994)
with relatively low diffuse X-ray emission (Snowden \& Petre 1994). The sightline towards Sk\,$-$67D05 
was the first LMC target investigated with {\small FUSE} showing large amount of molecular hydrogen (Friedman et al. 2000).
Fits of the Na\,{\small I} and Ca\,{\small II} lines reveal up to 12 components
in four distinct groups as shown in Fig. 3: the Galactic group centered at 0 km s$^{-1}$, an Intermediate Velocity Cloud (IVC) at $+$77 km s$^{-1}$,
and High Velocity Cloud (HVC) at $+$144 km s$^{-1}$, and the LMC group centered at $+$280 km s$^{-1}$.
Our {\small VLT} data show the hidden complexity
behind the simple picture derived at
lower resolution (15,000) by Friedman et al. (2000) using {\small FUSE} data alone.
A detailed summary of these Na\,{\small I} and Ca\,{\small II} components is given in Table 7.

The lack of Na\,{\small I} absorbers in between the IVC and HVC groups is quite remarkable for
an LMC target since large absorptions at $\approx$ $+$60 km s$^{-1}$ and $\approx$
$+$120 km s$^{-1}$ are usually observed along LMC sightlines. However, in their survey of
Magellanic sightlines with {\small IUE}, Savage \& de Boer (1981) reported the detection of
an IVC and a HVC clouds in the light path to Sk\,$-$67D05 through low ions absorptions
(Mg {\small I}, O {\small I}, Si {\small II}, and Fe {\small II}). 
Danforth et al. (2002) recently confirmed
the galactic origin of these absorbers from a deep investigation of dozens of {\small FUSE}
LMC sightlines. They demonstrated that both the IVC and the HVC are located in the general
direction of the Eastern edge of the LMC so that the Western line of sight towards Sk\,$-$67D05
is likely sampling only a small fraction of this material. 

Molecular hydrogen absorption lines are detected in the {\small FUSE} data.
Rotational levels up to $J=$ 3 were detected in the MW group and up to $J=$ 6 in the LMC group.
Because some of the molecular hydrogen absorption lines are expected to be saturated, it is mandatory to use as much information as possible
on the velocity structure of the molecular components. Although we do not have direct
information from molecular tracer species with our high resolution {\small VLT} data, the detailed
investigation of the Na\,{\small I} and the Ca\,{\small II} lines is used to
infer a plausible molecular velocity structure. Firstly, using the Na\,{\small I} lines
alone, we can constrain the position of the molecular absorbers to the position of the atomic
neutrals (as seen in Na\,{\small I}) and compare the model with the {\small FUSE} data. In any
case, we
shall expect to see at most the same number of molecular components as there are atomic
components. However, when the separation between component is less than 10 km s$^{-1}$ (well
below the {\small FUSE} resolution) the profile fitting of the {\small FUSE} spectra is less reliable .
Secondly, we can use the Na\,{\small I}/Ca\,{\small II} ratio to
diagnose which atomic component is cool and dense enough to harbor molecules. Typically,
the greater the ratio the more likely the presence of dust grains and then molecules (Welty et al. 1996).

Among the possible Na\,{\small I} MW absorbers, two were found to match the velocity of the {\bf {\small
FUSE}} H$_2$ absorption feature: one component at $+$18 km s$^{-1}$ and another component at $+$22 km s$^{-1}$. 
We then performed a two components analysis of the MW molecular content and find a total H$_2$
column density of $\log$ $N$(H$_2$) $=$ 15.24 $\pm$ 0.17 dex, summed over the $J=$ 0--3 states (Table 8), while the
rotational temperature $T_{01}$ is about 303 K. These properties are typical of diffuse
molecular clouds found in the halo as observed by Richter et al. (1999).
It should be noted that our column density estimates for the individual rotational
levels in the MW molecular component are somewhat larger than the estimates of Friedman
et al. (2000).

There are four atomic components spanning over 30 km s$^{-1}$ seen at the location of the LMC
velocities.
Imposing the 4-components solution into the {\small FUSE} data showed that the main LMC molecular
absorber
is associated with the atomic component at 291 km s$^{-1}$ while less than 2 \% of the H$_2$
is arising in the other components. We thus assume that the LMC molecular absorption lines detected
in
{\small FUSE} are from this single component. We find an H$_2$ column density of $\log$ $N$(H$_2$) $=$
19.44 $\pm$ 0.05 dex. With the exception of the $J=$ 3 measurement, all of our estimates of the molecular hydrogen rotational levels
in the LMC component are consistent with the survey by Tumlinson et al. (2002). We thus
derive similar temperatures with $T_{01}$ $\approx$ 57 K and  $T_{23}$ $\approx$ 225 K.
The population of the rotational levels might then be used to infer the incident radiation
field upon this LMC absorber.

In the following, we assume that the incident flux is responsible for the population
of the rotational levels $J>$ 2. This is consistent with the observation
of two excitation temperatures for the H$_2$ component. The two temperatures are
interpreted as the signatures of different processes acting together to
populate the H$_2$ molecule from the lower levels all the way up. The levels
$J=$ 0--2 are predominantly populated by collisional processes, with typical
excitation temperatures below 100 K. In the Galactic disk, the population of higher rotational
levels 
are explained by the conjugated effect of collisions, UV pumping, and formation processes
(Spitzer \& Zweibel 1974).
However in the LMC, the radiation field is likely the main source of
excitation of these higher rotational levels.
Thus from the balance of the $J=$ 4 level we can derive
a direct estimate for the incident flux using the method developed in Wright \&
Morton (1979). 

The basic equilibrium equation can be translated in the following manner:
\begin{center}
$p_4$ . ($N_0$+$N_2$+$N_4$) $=$ $N_{J=4}$.$A_4$ + $N_4$
\end{center}
where $N_2$ is the number of photo-excited H$_2$ molecules in $J=$ 2, $p_4$ is the
probability of photo-excited molecules to cascade into $J=$ 4 and $A_4$ is the rate
of radiative decay from $J$=4 to $J$=2.
The number of photo-excited H$_2$ molecules in a given rotational level is proportional
to the number of available photons multiplied by the probability of being absorbed:
\begin{displaymath}
N_4=\int_{J=4}{(\frac{U_\lambda c d\lambda}{h\nu}) \times
(\frac{I_{\lambda0}-I_\lambda}{I_{\lambda0}})}
\end{displaymath}
If we assume that $U_\lambda$/$h\nu$ is constant over the far {\bf ultraviolet} range and equals to
($U_\lambda$/$h\nu$)$_{1000}$,
then the previous equations transform into:
\begin{displaymath}
N_4 = (\frac{U_\lambda c}{h \nu})_{1000}\times\Sigma_{J=4}W_{\lambda}
\end{displaymath}
and
\begin{displaymath}
p_4.(\frac{U_\lambda c}{h \nu})_{1000}.\Sigma_{J=0,2,4}W_\lambda=N_{J=4}.A_4+
(\frac{U_\lambda c}{h \nu})_{1000}.\Sigma_{J=4}W_\lambda
\end{displaymath}
\noindent
with $p_4=$ 0.26 and $A_4=$ 2.76 $\times$ 10$^{-9}$ s$^{-1}$.
The calculations of the total equivalent widths ($\Sigma W_\lambda$) are then
computed from models of the different rotational levels based on our measurements.
Solving these equations yields $U_\lambda\approx$ 1.9 $\times$ 10$^{-16}$ erg cm$^{-3}$ \AA$^{-1}$ which
is about three times the assumed Galactic value of $\approx$ 7 $\times$ 10$^{-17}$
erg cm$^{-3}$ \AA$^{-1}$ (see Draine 1978; Parravano et al. 2003).

Then, with the incident flux we can estimate the distance of the cloud to the
background star. The large aperture of the {\small FUSE} satellite is sampling a
$30\arcsec\times30\arcsec$ sky area which represents about 8 pc $\times$
8 pc at the distance of the LMC. Since we do
not see any other hot stars in the {\small FUSE} aperture, we
assume that the background star is the main source of photons onto the
cloud. 
Thus, following Bluhm \& de Boer (2001), we calculate the distance using the simple
cross product $\frac{I_{\rm observed}}{I_{\rm cloud}}=(\frac{d}{r_{\rm LMC}})^2$. $I_{\rm observed}$
is the stellar continuum intensity at 1000 \AA~corrected for the extinction
using the model of Cardelli, Clayton \& Mathis (1989) $I_{\rm cloud}$ the intensity at the location
of the cloud derived from our calculation, $r_{\rm LMC}$ is the distance to the LMC ($\approx$
50 kpc) and $d$ is the distance between the cloud and the background star.
We find that the cloud cannot be any closer than 120 pc from Sk\,$-$67D05. This fact alone
suggests that the molecular component we are looking at is probably not associated with the
star formation region. The details the MCs H$_2$ investigations ($J$ levels, temperatures,
radiation fields and $b$-values) towards each sightline are shown in Table 9.

The derived values for the
HD/H$_2$ ratio and the CO/H$_2$ ratios are respectively 1.5$\pm0.5$ ppm
and 2.7$\pm0.7$ ppm. These values are comparable to the ones derived in the MW disk
(Ferlet et al. 2000; Lacour et al. 2003 in prep). Several important questions are raised by
these relatively large ratios. Firstly, the CO column density is directly influenced by
the availability of C and O atoms in the gas phase; while the C and O atoms are each
depleted by about 0.5 dex (Russell \& Dopita 1992), the CO seems not to be depleted
at all. Secondly, the lower
metallicity should also result in less dust grain surfaces and then in suppressed H$_2$
formation. But most of the CO lines
appear near the core of the H$_2$ absorption bands and it is usually assumed that
H$_2$ is an efficient photo--dissociation shield for CO.
The general expectation is then that the CO/H$_2$ ratio
would then tend to decrease when the metallicity decreases.
Thirdly, the relatively small H$_2$ column we derive should result
in a less efficient FUV--shield in this LMC cloud and a smaller CO/H$_2$ ratio whatever dust content there is along the sightline.

We also detected
neutral chlorine atoms which can only exist in the molecular component
(see section 3.5). We find Cl\,{\small I}/H$_2$ ratio of 2.74$\pm0.9$ ppm
a value at least twice higher than what has been found in the Galactic disk towards HD\,192639 (Sonnentrucker et al. 2002) and HD\,185418 (Sonnentrucker et al. 2003). Detailed analysis of the latter lines of sight indicated that the gas was made of multiple neutral diffuse clouds. The fact that we derive a much higher ratio along a sightline where metallicity is lower by 0.5 dex with respect to the Galactic values, strongly suggests that 
denser molecular clumps are present towards Sk\,$-$67D05. 

It is possible to get a crude estimate of the total H\,{\small I} column density in the LMC using the gas--to--dust ratio derived by Koorneef (1982) of $N$(H\,{\small I})/$E_{B-V}$
$=$ 2 $\times$ 10$^{22}$ mag$^{-1}$ cm$^{-2}$. However, we must correct the $E_{B-V}$ from potential MW gas contamination. Thus we calculated the
MW reddening assuming $N$(H$_2$)/$E_{B-V}$ $=$ 5 $\times$ 10$^{20}$ mag$^{-1}$ cm$^{-2}$ (Dufour 1982). We find 
that, towards Sk\,$-$67D05, the Galactic material represents only a small fraction
of the total reddening of about $E_{B-V {\rm (MW)}}$ $\approx$ 0.01, and thus a small fraction of the total
$N$(H\,{\small I}) of about 10$^{19}$ cm$^{-2}$. Consequently,
we derive $N$(H\,{\small I})$_{\rm total}$ $\approx$ $N$(H\,{\small I})$_{\rm LMC}$ $\approx$ 3 $\times$ 10$^{21}$ cm$^{-2}$.
This H\,{\small I} column density is about 6 times larger than the one derived towards
the same line of sight by Friedman et al. (2000). The difference is mainly due to the
fact that Friedman et al. (2000) used the galatic gas--to--dust ratio which is
4 four times smaller than the LMC value {(Koorneef et al. 1982)}. In our case, where
most of the H\,{\small I} is originating in the LMC, this can lead to a great systematic
error. Furthermore, we adopted the updated $E_{B-V}$ as quoted in Tumlinson et al. (2002)
which is 1.5 times greater than the value used by Friedman et al. (2000). 

It could be argued that Na\,{\small I} might as well be used as a fiducial
for H\,{\small I} with greater accuracy than the gas--to--dust ratio.
Indeed, investigations of
the column density correlation $N$(Na\,{\small I}) versus $N$(H\,{\small I}) by Ferlet et al.
 (1985) in the MW have demonstrated that Na\,{\small I} roughly follows H\,{\small I} over more than three orders of magnitude.
However the abundance of sodium in the LMC gas is varying between sightlines.
In the detailed investigation of Welty et al. (1999) towards SN1987A, about 20 individual LMC
components were found with $\log$ $N$(Na\,{\small I})/$N$(H$_{\rm total}$) ranging from $\approx$ $-$9.0  to $-$9.7 dex.
Here, we assume that the ratios between Na\,{\small I}
components are reasonably linked to the ratios between H\,{\small I} components and
we derived $N$(H\,{\small I}) $\approx$ 8 $\times$ 10$^{20}$ cm$^{-2}$ 
within the absorber associated with the H$_2$ component leading to a molecular
fraction $f$(H$_2$) $=$ 2\,$N$(H$_2$)/($N$(H{\small I})+2\,$N$(H$_2$)) around 7 \% on average over the Na components. This result together with the finding that, on average, the CO population ratio is subthermally distributed suggests that some more diffuse gas is also present along the Sk\,$-$67D05 light path.

Finally, we also measured the column density of Fe\,{\small II} and found a value consistent with the
analysis by
Friedman et al. (2000): $N$(Fe\,{\small II}) $=$ 7.0$^{+3.0}_{-1.5}$ 10$^{14}$ cm$^{-2}$. Note that this represents the
total amount
of Fe\,{\small II} as seen in the four LMC components. Although the profile fitting made use of
the four
components at the same time, interpretation of Fe\,{\small II} in each component is difficult
because these are not resolved by {\small FUSE}. Fortunately, most of the Fe\,{\small II} lines used in the profile
fitting are in the optically thin regime. Hence, the total Fe\,{\small II} we quote remains accurate. The
observed abundance
of Fe is then $\log$ [$N$(Fe\,{\small II})/$N$(H$_{\rm total}$)]$+$12 $=$ 5.4 where $N$(H$_{\rm total}$) is the total amount of
H atoms derived
from the four absorbers using Na\,{\small I}. When compared with the Fe abundance in
the LMC (Russell \& Dopita 1992), our result implies a depletion of about 2 dex. Hence large Fe\,{\small II} depletion occurs towards Sk\,$-$67D05 although it is not clear if Fe\,{\small II} is depleted onto grains or in the ionized form. The fact that the CO, HD and Cl\,{\small I} analyses strongly suggest the presence of dense molecular clumps could argue in favor of depletion onto grains but the present data does not allow us to rule out the possibility that Fe\,{\small II} might be predominantly ionized. 

\subsection{Sk\,$-$68D135}

Sk\,$-$68D135 is a ON9.7 star located in the North of 30 Doradus region which is the largest H
{\small II} region
in the local group of Galaxies (Fitzpatrick \& Savage 1984, Tumlinson et al. 2002). It is believed that this region
requires a Lyman continuum radiation field equivalent to some 100 O5 stars to maintain its
ionization. The H\,{\small I} data from Wayte (1990) at 8 km s$^{-1}$ resolution reveal the presence of at least four
absorbers along the sightline. Two of them arise in the
MW disk and halo around $+$30 and $+$90 km s$^{-1}$ in heliocentric velocity, while the two other are originating
in the LMC at $+$210 and $+$290 km s$^{-1}$ in heliocentric velocity. Our data however indicate the presence of 5 groups of absorbers split into 17 components and centered at (1) $+$30 km s$^{-1}$, (2) $+$80 km s$^{-1}$,
(3) $+$160 km s$^{-1}$, (4) $+$220 km s$^{-1}$, and (5) $+$280 km s$^{-1}$ (Table 7). The four components observed by Wayte (1990)
can then be identified with group 1, 2, 4, and 5. We thus confirm that these components are in the foreground of
Sk\,$-$68D135. We note also a LMC HVC located at $+$320 km s$^{-1}$ showing a large Ca\,{\small II} absorption relatively
to the corresponding Na\,{\small I} absorption and probably indicating shocked material.
Our profile fittings of the Ca\,{\small II} and Na\,{\small I} {\small VLT} data are shown in Fig.4.

H$_2$ arising in the MW is detected in the {\small FUSE} spectrum and we identify this molecular
absorber with the atomic absorber seen in Na\,{\small I} at 41 km s$^{-1}$. Not only is this
absorber the strongest in Na\,{\small I} but it also exhibits Na\,{\small I}/Ca\,{\small II} ratio
of the order of 0.3 dex, typical of cold components (Welty et al. 1999). Note that adopting the complete
Na\,{\small I} velocity structure does not improve the fit. This MW molecular component
is a lot larger than the one observed towards Sk\,$-$67D05. A summary of the MW H$_2$ analysis is given
in Table 8. We obtain $\log$ $N$(H$_2$) $=$ 18.29 $\pm$ 0.04 dex , $T_{01}$ $\approx$ 103 K, and $T_{23}$ $\approx$ 102 K.

The investigation of the velocity structure for the molecular species observed in the LMC is a
little
more difficult since there are four LMC components blended within 20 km s$^{-1}$. The profile
fitting
shows that the best solution is consistent with a dominant absorber at $+$269 km s$^{-1}$ along
with
marginal contribution from an absorber at $+$277 km s$^{-1}$ (less than 1 \%). We note that the absorber at $+$269 km s$^{-1}$ has the largest Na\,{\small I}/Ca\,{\small II} ratio of the four
components
and is the most likely to harbor dust grains and molecules.
After profile fitting analysis assuming a single absorber, we confirm most of the H$_2$
measurements
performed in the survey of H$_2$ Magellanic Clouds sightlines by Tumlinson et al. (2002).
However,
for the higher levels, our estimates of the mean column density are somehow different
and lead us to different excitation temperatures. We find $\log$ $N$(H$_2$) $=$ 19.87 $\pm$ 0.05 dex,
$T_{01}$ $\approx$ 92 K
and $T_{23}$ $\approx$ 251 K. Calculation of the incident radiation field
using the same assumptions as for Sk\,$-$67D05 gives $U_\lambda\approx$ 2.2$\times$ 10$^{-13}$ erg 
cm$^{-3}$ \AA$^{-1}$ which is several thousand times larger than the Galactic radiation field.
This is a tremendously high value suggesting that the cloud is located close to the star.
If we assume that the background star is the only source irradiating the cloud, we then
find that the cloud is located at 5 pc from the star. A likely scenario involves a larger molecular
cloud being disrupted around the star Sk\,$-$68D135.
We note that the incoming radiation field we derive is similar to the
ambient radiation field estimated in the 30 Dor region using CO line emissions
(Bolatto et al. 1999 derived $\chi$ $\approx$ 3000). Another hint for a large radiation
field comes from the observation of the strong Fe\,{\small III} lines at the LMC velocity.
Fig. 8 compares this line to the one observed towards Sk\,$-$67D05. Unfortunately,
because the line is heavily saturated it was not possible to give a quantitative measurement
for the Fe\,{\small III}/Fe\,{\small II} ratio. We note finally that this
intense radiation field suggests that the cloud is embedded within the 30 Dor region
and, therefore, may be at any distance from the background star.

That the CO and HD molecules are detected in such a harsh environment is puzzling. We actually find CO/H$_2$ ratio of 0.80 ppm and a HD/H$_2$ ratio of 1.9 ppm, values
more typical of translucent ($A_V$ $>1$) Galactic sightlines (Rachford et al. 2002).
Despite the low chlorine LMC abundance (depletion of 0.5 dex, see 
Russell \& Dopita 1992), we find that the Cl\,{\small I}/H$_2$ ratio is about 0.5 ppm again comparable to the Galactic measurements. These results, as for Sk\,$-$67D05, strongly suggest that higher density molecular clumps exist toward Sk\,$-$68D135 too.

Using the LMC gas--to--dust ratio (see section 4.1), we find that
the LMC H\,{\small I} column density can be as high as 5 $\times$ 10$^{21}$ cm$^{-2}$.
To derive the fraction of the molecular absorbers we assumed that the
Na\,{\small I} ratios between components translate into H\,{\small I} which would lead
to a H\,{\small I} column density of about 3 $\times$ 10$^{21}$ cm$^{-2}$ and a molecular fraction
of 1 \%. Models, however, show that such a low molecular fraction cannot
be reconciliated with the detected amount of HD molecules. The photodissociation time scale
for the HD molecule is given by the photodissociation rate (2.6 $\times$ 10$^{-11}$ s$^{-1}$
in the Galaxy, see Le Petit et al. 2002), $T_{\rm phot}$ $\approx$ 1,500 years. And yet
this time scale should be smaller in the LMC where more photons are available.
When compared to the age of the star, $T_{\rm star}$ $\approx$ 10$^6$ years, this time-scale
is ridiculously small and suggests that HD is locked in a dense phase where the molecules
are efficiently self-shielded and where the formation routes are enhanced.

As a matter of fact, because of the location of the cloud within the 30 Dor region, and if the cloud is a remnant of a denser molecular knot, we expect the survival of CO, HD and
other molecules in the core while the envelope, and all the H$_2$ within it,
is being swept away by the shocks or the stellar winds or destroyed by photons (thus
reducing the measured integrated molecular fraction), as is observed.
We note also that strong shocks are in action in 30 Dor and
some of the dust grains might be destroyed, releasing in the gas phase many metals.
The presence of a shock along the sightline to Sk\,$-$68D135 is reinforced by the
detection of the LMC HVC component at $+$308 km s$^{-1}$ with an extremely low
Na\,{\small I}/Ca\,{\small II} ratio of $-$0.5 dex.

\subsection{Sk\,$-$69D246}

The sightline towards Sk\,$-$69D246 has been extensively studied by Bluhm \& de Boer
(2001). Sk\,$-$69D246 is a WN6 star located nearby R136, at the edge of the 30 Dor complex. 
{\small IUE} absorption lines of S\,{\small II}, and Mg\,{\small I} as well as {\small FUSE}
Fe\,{\small II} absorption lines show a velocity structure typical of most of the LMC
sightlines (Danforth et al. 2002). We find three MW components around 0 km s$^{-1}$, a disk cloud at
$+$40 km s$^{-1}$, one IVC at $+$80 km s$^{-1}$, one HVC at $+$130 km s$^{-1}$, and four LMC
components at $+$234 km s$^{-1}$, $+$260 km s$^{-1}$, $+$280 km s$^{-1}$, and $+$290 km s$^{-1}$.
Our high resolution {\small VLT} data of Na\,{\small I} and Ca\,{\small II} make it possible to break the
degeneracy associated with blended components not seen in {\small IUE} (Fig. 5). We are indeed able to detect
18 components as described in Table 7.

The MW molecular component is associated with the component at 25 km s$^{-1}$ (as seen in
Na\,{\small I}) and has the same properties as the one seen towards Sk\,$-$68D135
except for a slightly larger total H$_2$ column density. It is likely that both sightlines are
crossing the same Galactic molecular cloud and the comparison could bring interesting limits on
the variation scale across this absorber (see Sect 5.4). We find $\log$ $N$(H$_2$) $=$ 18.73 $\pm$ 0.04 dex and calculations
of the Boltzamn temperatures yield $T_{01}$ $\approx$ 83 K and $T_{23}$ $\approx$ 179 K.

The column densities for $J=$ 0 to $J=$ 6 in the LMC molecular component at $+$279 km s$^{-1}$ were
derived by profile
fitting and we noted that our fits improved noticeably when adding a second molecular component
at $+$285 km s$^{-1}$
in heliocentric velocity.
{\small FUSE} cannot fully resolve the two components, but
we point out that a similar velocity structure was derived by Bluhm \& de Boer (2001).
Furthermore, Cl\,{\small I} {\small STIS} data clearly confirm the presence of two
molecular
absorbers as shown in Fig. 5. For the total LMC H$_2$ column densities we find $\log$ $N$($H_2$) $=$ 19.66
$\pm$ 0.04 dex , $T_{01}$ $\approx$
 70 K and $T_{23}$ $\approx$ 352 K. The balance of the $J=$ 0, 2 and 4 levels yields (see section
4.1 for details)
$U_\lambda\approx$ 1.6 10$^{-15}$ erg cm$^{-3}$ \AA $^{-1}$. Assuming that the high J levels are
mainly populated by UV--pumping and assuming that the star is the principal source of UV photons, the clouds cannot
be closer than 28 pc from the background star Sk\,$-$69D246 and are possibly embedded in the 30 Dor region.

CO, HD and Cl\,{\small I} are also detected in the LMC molecular components in relatively large
amounts. The CO/H$_2$ ratio is of the order of 0.8 ppm, the HD/H$_2$ ratio
is 0.9 ppm and the Cl\,{\small I}/H$_2$ ratio is 0.6 ppm. {\it A priori}, these values are clearly
unexpected in the LMC, especially in 30 Dor unless, as for the two other stars, dense clumps exist and shield these molecules from the high UV radiation field. 

We derive the H\,{\small I} column density associated with the LMC molecular components following
the two assumptions described in section 4.1:
$N$(H\,{\small I}) $=$ 2 10$^{21}$ cm$^{-2}$, two times larger
than the one quoted in Bluhm \& de Boer (2001) who used a direct relation between S\,{\small II} and H
{\small I}.
Because the two S\,{\small II} components are not resolved in the {\small IUE} data and because
the S\,{\small II}
lines are saturated, the analysis of S\,{\small II} data is difficult and does not allow to
take into account
saturation effects in the core of the absorption lines.
Thus, we cannot discard the fact that the H\,{\small I} column density may be underestimated in the work of
Bluhm \& de Boer (2001).
The iron abundance is 5.96 leading to a depletion of
1.3 dex compared to the assumed LMC abundance (Russell \& Dopita 1992). Iron seems therefore to be strongly depleted. At this point, we cannot decide whether iron was preferentially depleted onto grains or strongly ionized by the high ambient radiation field. 

With the relatively high signal-to-noise of the {\small FUSE} spectrum, the LMC Cl\,{\small II} absorption
line is clearly detected at 1071 \AA. At the same time, higher
resolution {\small STIS} data of the Cl\,{\small I} line at 1347 \AA\ are also available. We have seen that 
Cl\,{\small II} is expected in the diffuse ISM where the hydrogen
is atomic while Cl\,{\small I} is associated with the molecular ISM.
Adopting the Jura \& York (1978) model of a single cloud made of a skin,
mainly atomic, and a core, mainly molecular, one can estimate the column density $N_1$(H) of H\,{\small I} atoms co--located with the optically thick H$_2$ gas as follows: $N_1$(H) $=$ $N$(H$_{\rm total}$)($f$(Cl\,{\small I}) $-$ $f$(H$_2$)) where
$f$(Cl\,{\small I}) is the fraction of neutrals (Cl\,{\small I}) among the chlorine atoms, $f$(H$_2$) is the integrated
molecular fraction along the sightline, $N$(H$_{\rm total}$) is the integrated total column of hydrogen atoms  (atomic and
molecular). We
estimate that the molecular fraction inside the core of the molecular component, assuming
similar physical properties and chlorine depletions, is about 12\%. It should be noted
that this is a strong lower limit since some of the diffuse H\,{\small I}
is possibly arising in separate pure atomic components and thus $N_1$(H) is possibly
overestimated.

The LMC C\,{\small I}, C\,{\small I*}, and C\,{\small I**} lines are also detected towards 
Sk\,$-$69D246 as shown in Fig. 9. Although many absorption line multiplets are present in the {\small FUSE} range, it is
difficult to isolate unblended lines. We therefore employed the {\small STIS} data around 1275
\AA\ to perform the analysis as shown in Fig. 7, when possible. These lines were assumed to arise only in the
molecular components and were fitted along with the LMC molecular hydrogen (same velocity structure,
same $b$-values). The profile fitting analysis yields $\log$ $N$(C\,{\small I}) $=$ 13.85$^{+0.06}_{-0.06}$,
$\log$ $N$(C\,{\small I*}) $=$ 13.72$^{+0.10}_{-0.10}$, and $\log$ $N$(C\,{\small I**})
$=$ 13.44$^{+0.05}_{-0.06}$ (see Table 6). The total fractional
abundances of carbon in the excited states are $f_1$ $=$ $N$(C\,{\small I*})/$N$(C{\small I}$_{\rm total}$) $=$ 0.43$\pm$0.08 and $f_2$ $=$
$N$(C\,{\small I**})/$N$(C\,{\small I}$_{\rm total}$) $=$ 0.16$\pm$0.02. Assuming an average radiation field
of ten times the Galactic radiation field, we obtain $\log$ ($P/k$) $=$ $\log$($n_{\rm H}T$) between 4.3 and 4.6
(Jenkins \& Shaya 1979). If we assume that
the carbon atoms are in the cold LMC molecular component at 72 K, then we can expect the
density $n_{\rm H}$ to be roughly in between 300 and 600 cm$^{-3}$. This should
be considered as a preliminary estimate since, among other simplifications, we do not take into account the multiplicity of the sightline.

Additional evidence towards the existence of a dense component comes from the detection of the
$\lambda$4227 \AA\ Ca\,{\small I} line at the LMC velocity. Even if the velocity structure of the
line
is buried within the noise of the spectrum, we can safely derive the total column density since
the line
is optically thin. We find $N$(Ca\,{\small I}) $=$ 1.5$^{+0.2}_{-0.2}$ 10$^{10}$ cm$^{-2}$. If we assume that the ionization
equilibrium pertains, the Ca\,{\small I}/Ca{\small II} ratio
is a function of $n_e$, the radiation field and the temperature via the recombination coefficient $\alpha$
and the photoionization rate $\Gamma$ (see section 3.6). Using $\alpha$ $=$ 5.58 $\times$ 10$^{-12}$ cm$^3$ s$^{-1}$ at T
$\approx$ 100 K
and $\Gamma$ $=$ 37 $\times$ 10$^{-11}$ s$^{-1}$ (see P\'equignot \& Aldrovandri 1986) multiplied by 20 to account for
the larger radiation field, we find $n_e$ ranging from 0.1 to 1 cm$^{-3}$. The latter values are obtained if we add the Ca\,{\small II} component observed
at 295 km s$^{-1}$
which is probably too warm to be associated with the molecular absorbers. Compared to the
previous density estimate, this
electronic density leads to an ionization fraction of 1 \% at most.

\subsection{AV\,95}

With the exception of the O\,{\small VI} lines studied by Howk et al. (2002),
no investigation of the ISM towards AV\,95 was found in the literature. Combining {\small FUSE}
observations of Fe\,{\small II} with {\small STIS} observations of S\,{\small II} we detect 11
atomic components. Thus, this sightline is quite typical of SMC sightlines (Welty et al.
1997; Mallouris et al. 2001) showing a number of low-ionization Galactic gas clouds
or complexes spread over $\approx$ 65 km s$^{-1}$ (Fig. 10). Note that Fe\,{\small II} and S\,{\small II}
ions are tracing the neutral component as well as the H\,{\small II}
regions. Both species seem to share the same velocity structure but exhibit
different ratios between components. Fe\,{\small II} in particular exhibits a few extra high velocity components at $+$169 km s$^{-1}$
and $+$195 km s$^{-1}$ (heliocentric velocities) that could correspond to shocked gas in which the grain sputtering released some of the refractory elements such as Fe into the SMC gas phase. The detailed velocity structure is summarized in Table 10.

The profile fitting of the {\small FUSE} data along with available {\small STIS}
observations of Cl\,{\small I} at $\approx$ 6 km s$^{-1}$ resolution is consistent with the presence
of a single absorber in the MW and in the SMC with respective $b$-values of 1.5 km s$^{-1}$ and
0.8 km s$^{-1}$. For the MW component we derive $T_{01}$ $\approx$ 66 K, $T_{23}$ $\approx$ 135
K, $\log$ $N$(H$_2$) $=$ 18.26 $\pm$ 0.05, and $N$(Cl\,{\small I}) $=$ 8.0 10$^{12}$ cm$^{-2}$
which leads to a Cl\,{\small I}/H$_2$ ratio of 0.4 ppm.
Despite the large H$_2$ column density we do not detect the HD molecule in this MW component.

The SMC C\,{\small I}, C\,{\small I}*, and C\,{\small I}** lines at 1276.48 \AA,
1276.75 \AA, and 1277.72 \AA, available in the {\small STIS} spectra (Table 6), were first fitted
along with SMC molecular hydrogen assuming common velocity structures and $b$-values.
However, we found that these two assumptions were too strong and did not allow a
proper fit of the carbon lines. Indeed, it appears
that some of the carbon absorption lines arise in several components. Because our data did
not allow us to reliably trace these more diffuse components, we performed a one-component fit
that led to a total $b$-value of 5.7 km s$^{-1}$ (about 6 times greater than the SMC H$_2$ adopted
$b$-value). We note, that from the S\,{\small II} absorption lines detected in the {\small FUSE} range,
there are many atomic diffuse components on this pathlength and some may well be cool enough
to exhibit C\,{\small I}, C\,{\small I*}, and C\,{\small I**} lines. With these carbon lines, we inferred a range of possible H number densities for this component between 100 to 1000 cm$^{-3}$.
Due to the SMC blending it is not possible to derive the H\,{\small I} content directly. One alternative way is
to estimate
the total sulfur column density and calculate the total hydrogen column density assuming the
SMC standard abundance ($\log$ $N$(S\,{\small II})/$N$(H\,{\small I}) $=$ 7.27; see Russel \& Dopita 1992) and no
depletion into
the grains.
The total Galactic sulfur column density is hence $N$(S) $=$ 2 $\times$ 10$^{15}$ cm$^{-2}$ (the dominant ionization stage is S\,{\small II}) which
leads to $N$(H$_{\rm total}$) of the order of 10$^{20}$ cm$^{-2}$, and $f$(H$_2$) of only 3 \%. The high HD/H$_2$ and Cl\,{\small I}/H$_2$ ratios therefore suggest that denser clumps exist and are mixed with the more diffuse gas also present along the light path to AV\,95 in the SMC.

The SMC molecular hydrogen is similar to the LMC ones with $T_{01}$ $\approx$ 78 K,
$T_{23}$ $\approx$ 245 K,
and $\log$ $N$(H$_2$) $=$ 19.43 $\pm$ 0.04. As above, we derive the H\,{\small I} column density, using S\,{\small II} ($\log$ $N$(S\,{\small II})/$N$(H\,{\small I})$_{\rm SMC}$ $=$ 6.6; see Russel \& Dopita 1992) and find $N$(H\,{\small I})
$\approx$ 10$^{21}$ cm$^{-2}$ and $f$(H$_2$) $\approx$ 2.5 \%. Again, some diffuse gas must be
laying
along the path length in the SMC gas too. From the population of the $J=$ 0, 2 and 4 levels we derive a radiation
field
af 1.8 $\times$ 10$^{-14}$ erg cm$^{-3}$ \AA$^{-1}$ which is 250 times greater than the average Galactic
radiation field.
Hence, assuming that most of this incident flux is coming from the background star, we get
a lower limit
on the distance between the star and the absorber of 3 pc.

The SMC CO molecule is barely detected in the
{\small FUSE} spectra. We report a stringent 3 $\sigma$ upper limit of $\log$ $N$(CO) $<$ 13.23.
This result is consistent with the fact that fewer oxygen atoms and fewer carbon atoms are
available to form the CO molecule and that CO is more photodissociated toward this line of sight the molecular clump
density being probably lower (see our previous C\,{\small I} results).

The SMC Cl\,{\small I} line is clearly detected towards this star with the {\small STIS} data
 (1347 \AA). The Cl\,{\small I}/H$_2$ ratio is 0.4 ppm  and translates into a depletion as low as
0.70 dex as shown in Table 9, assuming [Cl] $=$ 4.70 in the SMC (Russell \& Dopita 1992).
However, we note that the possible presence of Cl\,{\small II} might explain partly the low value. Indeed, the lower limit derived for the Cl\,{\small II} ion is consistent with
no depletion of chlorine atoms at all. Another test of the depletion along the sight line is
given by the Fe\,{\small II} atoms. Using H\,{\small I} derived from S\,{\small II}, we find an iron
abundance of $\approx$ 6.4 instead of 6.84 as derived in the SMC by Russel \& Dopita (1992).
The relatively low iron depletion, only 0.44 dex instead of 2 dex usually reported in the
MW ISM (Savage \& Sembach 1996), can be due to a low dust content or is the signature of shocks,
past or present (as revealed by the two SMC HVCs at $+$169 km s$^{-1}$ and $+$195 km s$^{-1}$).
Even though the depletion is uncertain, there are clear hints of a cold medium ($T_{01}$ = 78 K) with a
$b$-value as low as 0.8 km s$^{-1}$ (as derived from the simultaneous fit of several species) so that we
cannot rule out the presence of a dense core which might harbors a D--reservoir. A D--reservoir is defined as
the inner region of a cloud where all of
the deuterium atoms (hydrogen atoms) are in the form of HD (H$_2$) molecules and where the D/H ratio thus translates
into HD/2$\times$H$_2$; these hypothetical places are of cosmological interest since they would allow accurate
D/H measurements throughout the MW disk (Ferlet et al. 2000). The predominance
of diffuse gas conditions however renders the study of this colder denser molecular gas more difficult.

Relatively large amounts of the HD molecule are detected in the SMC towards AV\,95. Using both the AOD and PF techniques, we find $\log$ $N$(HD) $=$ 13.82$^{+0.96}_{-0.18}$.
The large error bar is due to the possible saturation of some of the HD lines detected  in the spectra.
Interestingly enough, HD/2 H$_2$ ranges from 1.6 ppm to 21 ppm within 1 $\sigma$. 
The latter value, if confirmed, would correspond to the first clear detection of a D--reservoir ever ({\bf Lacour} et al. 2004).
Moreover, the existence of a D--reservoir is not excluded by the range of densities derived, although crudely,
through the C\,{\small I} excited states. 

\subsection{Sk\,159}

Sk\,159 is located within the Knot 1 (K1) which is rich in H\,{\small II} regions
and is located in foreground compared with the SMC main body (De Vaucouleurs \& Freeman, 1972).
This region is remarkable for the homogeneity of the stellar radial velocities with a
typical dispersion of 5 km s$^{-1}$.
The investigation of this region based on Na\,{\small I} lines by Silvy (1996)
showed that Sk\,159 is located in the extreme edge of the H complex within the
K1 region. The latter data are consistent with the presence of a dominant SMC atomic absorber at $+$144
km s$^{-1}$ and a smaller one at $+$149 km s$^{-1}$.
In the {\small FUSE} data no intermediate components and no SMC HVCs are detected in Fe
{\small II}. The profile fitting in the {\small FUSE} range is consistent with a single atomic
component near $+$150 km s$^{-1}$. However, high resolution Na\,{\small I} data obtained at
the ESO 3.6m/CES/VLC by D. Welty (private communication) clearly show two identical
components separated by approximately 5 km s$^{-1}$. 
The relative simplicity of this sightline is likely the double effect of
a position of the star near the edge of the SMC and a weak foreground Galatic material.
The H\,{\small I} radio data obtained by Wayte (1990) show two components at $+$150 km s$^{-1}$
and $+$190 km s$^{-1}$. The non-detection of the H component at $+$190 km s$^{-1}$ in absorption
(Fig. 11) shows that most of the material seen in radio is located behind the star.

The MW molecular component is quite small with $\log$ $N$(H$_2$) $=$ 16.31 $\pm$ 0.25 and quite
warm with $T_{01}$ $\approx$ 153 K. Hence, this component ressembles those observed in the
Galactic
halo by Richter et al. (1999).

The SMC H$_2$ was fitted assuming a velocity structure similar to that of Na\,{\small I}
(see D. Welty, private communication). We thus performed profile fitting of the {\small FUSE}
data with this structure which led us to tentitatively detect 2 molecular components: one at $+$143 km s$^{-1}$ and
the other one at $+$152 km s$^{-1}$.
We find that the integrated properties are similar to the ones observed towards AV\,95
with $T_{01}$ $\approx$ 87 K, $T_{23}$ $\approx$ 204 K, and $\log$ $N$(H$_2$) $=$ 19.21 $\pm$ 0.06.
Calculation of the incident radiation
field gives 3.0 erg cm$^{-3}$ \AA$^{-1}$ which indicates that the clouds are at least at 50 pc from
Sk 159. We report
on a possible weak detection of the SMC CO molecule with $\log$ $N$(CO) $<$ 13.34 (3 $\sigma$ upper
limit).
On the contrary, the SMC HD molecule is clearly detected in the data with $\log$ $N$(HD) $=$ 13.85$^{+0.11}_{-0.14}$ and HD/H$_2$ $=$ 4.5 ppm.
Thus, we obtain a lower limit to the D/H ratio of 2.3 $\times$ 10$^{-6}$ inside that
component. The D--reservoir does not seem to be reached. We would like to point out
that this analysis is only based on {\small FUSE} data and a $b$-value of 2.6 km s$^{-1}$ but
higher
resolution data might unravel colder components buried in the absorption lines and then might
lead to a higher HD column density.

\section{Discussion}

\subsection{General Properties of the ISM in the MCs}

	In their large survey of the diffuse molecular ISM in the MCs, Tumlinson et al.
	(2002) have explored some of the global properties of this gas, comparing
	these properties to typical galactic properties, and derived
	a reduced H$_2$ formation rate, $R$ $\approx$ 3 $\times$ 10$^{-18}$ cm$^3$ s$^{-1}$ (one--third to one--tenth
	the Galactic rate),
	as well as reduced molecular fractions $f$(H$_2$) of the order of 1 \% ({\bf one--tenth}
	the Galactic mean). These results
	were interpreted as the effect of suppressed H$_2$ formation on grains and
	enhanced destruction by FUV photons. The total diffuse H$_2$ masses were
	then found to be of less than 2 \% of the H\,{\small I} mass suggesting either a
	small molecular content and a high star formation efficiency or the
	presence of substantial mass in cold, dense clouds unseen in their survey.
	{\bf Although we} have only 5 stars at our disposal, we can pursue the
	investigation of this ``pseudo'' diffuse molecular ISM of the MCs with
	a closer look at species not investigated previously.

        In this work, we have investigated three lines of sight towards the Large Magellanic Clouds and reported
        measurements of HD, CO, Cl\,{\small I} and H$_2$. A summary of these measurements is given in Table 9. The average
        HD/H$_2$ ratio is 1.4 $\pm$ 0.5 ppm, the CO/H$_2$ average ratio is 1.4 $\pm$ 0.5 ppm (we find 0.8 ppm
	when the large value towards Sk\,$-$67D05 is excluded), and the average
        Cl\,{\small I}/H$_2$ ratio is 1.3 $\pm$ 0.5 ppm where the quoted errors are the standard
        deviations. These measurements lead us to two major observations. Firstly,
	despite the large variation of physical conditions between the sightlines
	(the incident radiation field varies by 3 orders of magnitude, the total extinction varies
	by a factor 3 as well as the total H column densities) the standard deviation of these
	ratios is only of the order of 30 \%. Thus, it seems that each time the HD molecule is detected
	in the LMC ISM we are indeed sampling the same kind of isolated cloud whose individual properties
	are not averaged along the pathlength, and within which the
	shielding is efficent enough to smooth the influence of the radiation field.
	
Secondly, these ratios are similar to the ones derived in the Galactic ISM.
        This is a bit surprising given the low total extinction towards these stars,
        the important incident radiation field as derived from the UV pumping, and
        the low abundance of carbon and chlorine atoms in the LMC.
	The sightline towards Sk\,$-$68D135 gives the most extreme example. We have
	shown ($\S$ 4.2) that the  cloud is embedded within the 30 Dor region where
	the radiation field is about 3,000 times greater than the typical Galactic value.
	Contrarily to the expectations, this sight line exhibits the largest CO/H$_2$,
	the largest Cl {\small I}depletion, and the largest HD column density in the present sample.

        Towards the SMC, we report two detections of the HD molecule, one detection of Cl\,{\small I},
        and two upper limits on the CO molecules. The Cl\,{\small I}/H$_2$ ratios is 0.4 ppm towards AV\,95 well
        within the range of typical Galactic translucent clouds. The two lower limits that we derived  for the CO/H$_2$
	ratio are consistent with the fact that the CO abundance is
        expected to be significantly smaller in the SMC (see $\S$ 5.1). The measurement of the HD molecule towards AV\,95
	allows for a surprisingly high HD/H$_2$ ratio of 21 ppm (the largest value of the whole sample).
	As for CO, {\bf such a large amount} of HD molecules is unexpected along diffuse MC's line of sight
	and this observation strongly suggests that denser clumps are
	present and contribute significantly to the total column densities.

	However, such measurements may suffer from unknow systematics and
	we would like to state that the profile fitting of {\small STIS} Cl\,{\small I} data that strongly constrains the
	molecular components only shows saturated absorption lines for which the column densities are driven by the $b$-value.
        Our simultaneous analysis of {\small FUSE} and {\small STIS} data combining several species, however, confirms the presence of an extremely narrow SMC component (0.8 km s$^{-1}$ of total $b$-value) adding weight to the suggestion that denser molecular gas is present.

	Another interesting comparison of the MW  ISM with the MCs ISM
	comes from the molecular gas--to--dust ratio.
        With the direct measurements of the Galactic H$_2$ column densities towards the MCs,
        we are able to correct for the reddening from the foreground Galactic material (Dufour et al. 1982;
	see $\S$ 4.2) and derive the intrinsic MCs's reddening. Noteworthy is that for the five sightlines
	the corrections were found to be small which is consistent with the fact that our sample is
	biased towards low foreground Galactic column densities (see $\S$ 3.3). 
        Finally, we find that the H$_2$/$E_{B-V}$ in the LMC ranges from 1.7 to 5.2 $\times$ 10$^{20}$ mag$^{-1}$ cm$^{-2}$
	in Sk\,$-$67D05 and Sk\,$-$69D246 respectively. In the SMC we derive 4.7 $\times$ 10$^{20}$ mag$^{-1}$ cm$^{-2}$
	for the AV\,95 line of sight and 1.7 $\times$ 10$^{20}$ mag$^{-1}$ cm$^{-2}$ for Sk\,159.
        Contrarily to the conclusion drawn in Bluhm \& de Boer (2001) towards Sk\,$-$69D246, we find that these ratios
        compare quite well with the Galactic value from Dufour et al. (1982) of H$_2$/$E_{B-V}$ $\approx$
        5 $\times$ 10$^{20}$ mag$^{-1}$ cm$^{-2}$ but our small sample does not allow us to make further quantitative investigations.
        We wish to point out that the LMC value for the $N$(H\,{\small I})/$E_{B-V}$ ratio
        as derived by Koorneef (1982) is $\approx$ 2 $\times$ 10$^{22}$ mag$^{-1}$ cm$^{-2}$ and is several times larger than in
        the MW disk. This is somehow consistent with smaller molecular fractions in the LMC and the fact that
        less dust grains (meaning less reddening) are sampled in average towards any sightline
	in the LMC compared to the MW. 

	Our investigation of the LMC and SMC ISM leads us to propose the
	existence of individual dense molecular structures where the translucent properties
	of the studied sightlines arise. This would explain the remarkable constancy of
	the ratios we derived despite the drastically different galactic environments.
	Furthermore, we propose that these clouds should be dense enough to shield efficiently
	CO and HD despite enormous differences noted in the incident radiation fields.
	A preliminary estimate of the H number density using the study of atomic carbon fine structure lines
	observed with {\small STIS} towards Sk\,$-$69D246 and AV\,95 (see Table 6) led to estimate that $n_{\rm H}$ $>$ few 100 cm$^{-3}$.

	Finally, our analysis indicates that, as for the MW, the MC sightlines exhibit a mixture of diffuse and denser molecular gas. The cold molecular gas
	in the MCs is however more readily detected in the MC gas probably due to a lower diffuse/dense gas ratio. We indeed observe one order of
	magnitude less H$_2$ as towards typical Galactic translucent sightlines.  

\subsection{Molecular clumps or filaments in the MCs}
     
        It has been known for some time that the interstellar H {\small I} of the LMC
	is a cooler mixture of phases than in the MW. The first direct evidences were reported
	by Dickey et al. (1994) in their {\small ATCA} 21--cm survey within the LMC; they established that a
	cold atomic phase is present in the MC ISM at variance with the lack of cold phases
	outside of the optical boundary of the MCs (Mebold et al. 1991). Moreover, they showed that
	the cold atomic phase is more abundant in the LMC ISM than within the MW. A few years
	before, observations of the low CO luminosity of the MCs (Cohen et al. 1988; Israel et al. 1991)
	have been reported to be marginally consistent with the relatively large amount of atomic gas present
	($\approx$ 2 10$^9$ M$_\odot$). The difference between observations of cold atomic phases
	and cold molecular phases was soon interpreted to be an effect of the phase mixing and clumpiness
	of the medium (Lequeux et al. 1994).

	However, to date only a few hints towards a clumpy molecular phase are available.
	Recent ESO-SEST observations have revealed a new picture of the LMC molecular clouds
        with higher complexity than previously observed and molecular clumps as small
        as 25 pc or less, the limitation being the spatial resolution of the instrument.
        The SEST survey towards the LMC complex N11 by Israel et al. (2003)
        allows the identification of a population of dense molecular clouds ($n_{\rm H_2}$ $\approx$
        3000 cm$^{-3}$) with relatively warm temperature (60K--150K). This observation confirms the
        model proposed by Lequeux et al. (1994) in which the molecular complexes are seen as a collection
        of identical clumps of uniform density (between 10$^{3}$--10$^{5}$ cm$^{-3}$) immersed  in a non uniform
        interclump medium with density $n_{\rm H}$ $\approx$ 100 cm$^{-3}$.

        The detection of cold atomic clouds have also been reported by Kobulnicky \& Dickey (1999)
        in absorption in the Magellanic Bridge, a 10--degree region of diffuse gas linking the main
	SMC body to an extended arm of the LMC. Shortly after, Lehner (2002) confirmed the existence of
        small molecular absorbers in the Magellanic Bridge using {\small FUSE} data.
	Typical densities for diffuse molecular absorbers in the MW disk (Rachford et al. 2002;
	Sonnentrucker et al. 2002; Sonnentrucker et al. 2003) vary roughly in the
	range 1--100 cm$^{-3}$; assuming that these densities hold for the Magellanic bridge 
	and given the low column density reported by Lehner (2002) ($\log$ $N$(H$_2$) $<$ 15 dex)
        we compute the sizes of these clumps in the range 10--1000 Astronomical Units (AU).

        Our present observations of molecular clumps in the MCs are consistent with these later results:
	with typical
        $\log$ $N$(H$_2$) $\approx$ 19 dex and density $>$ 100 cm$^{-3}$, we derive sizes of 10$^4$
	AU or 1/10 pc (well below the 25 pc reported by Israel et al. 2003).

	The detection of CO molecules in our LMC sightlines is a strong evidence for even denser clumps
	for it is usually assumed that densities ranging from 1,000 to 100,000 cm$^{-3}$ are necessary to fit the
	observations (Lequeux et al. 1994). Recent investigations of the H$_3^+$ abundances led
	Cecchi-Pestellini \& Dalgarno (2000) to the same conclusions toward the Galactic star Cygnus OB2 no.12.
	Our observations favor as well the existence of dense clumps embedded in a tenuous interclump medium. Not only do we find that
	the CO/H$_2$ ratio is relatively high, indicative of denser gas, but also that the investigation of
	the rotational levels of the CO molecules in the LMC targets leads to $\langle$ $T_{ex}$(CO) $\rangle$
	$\approx$ 9 K typical
	of subthermal excitation occuring in cold clouds (see Fig 12).


	In fact, if we adopt the model of clumpy diffuse ISM derived by Cecchi--Pestellini
	\& Dalgarno (2002) we find densities of the order of 10$^4$ cm$^{-3}$.
	At face value, this leads to extremely small clump with  sizes of 10$^3$ AU, 10$^2$ AU, and smaller
	(the solar system is about 150 AU).
	However, several objections can be raised against the existence of such small molecular blobs.
	First, we can compare the probability for such an absorber to cross the light path to a
	MCs bright star with the actual covering factor.
	Assuming that the total molecular phase hidden in these blobs is of the order of
	10$^7$ M$_\odot$ (10$^9$ M$_\odot$ in cold atomic phase $\times$ a molecular fraction of 1\%,
	see Dickey et al. 1999 and Tumlinson et al. 2002) over a 1kpc $\times$ 1 kpc
	sky area (dimensions of the 30 Dor region), assuming that the blobs are randomly distributed in volume, then,
	knowing the radius (0.05 pc) and the mass of each blob ($\approx$0.1 M$_\odot$
	with $n_{\rm H}$ = 10$^4$ cm$^{-3}$) we infer a crude covering factor of the order of 40 \%.
	This is inconsistent with our observation that 3 LMC sightlines out of 55 ($\approx$ 5 \%)
	and 2 out of 26 SMC sightlines ($\approx$ 8 \%) are crossing these clumps (even taking into
	account a handful of eventual misdetections; see $\S$ 2.1). It should be noted,
	however, that in the above picture, it is assumed that most of the molecular mass resides in cold,
	dense clumps which is not demonstrated.
	Second, if we assume that these blobs are roughly spherical then the dimensions we derive
	are ill--defined or should be considered only lower limits to the total radius.
	In effect, there is a larger probability to cross the outskirts than the core of such a
	clump and the frequency of core--crossing sightlines remains, by nature, a hidden parameter.
	Anyhow, this later scenario fails also to predict the observed similarities between sightlines
	(for instance, all the absorbers have total H$_2$ column densities within a factor of two).

	Filementary structures are, then, an interesting alternative. A small fraction of the molecular
	gas condensed in such structures could result in a relatively large covering factor if
	they are preferentially found nearby the stars (this is the case for all our targets except for Sk\,--67D05).
	At the same time they provide a natural explanation for the
	about constant total H$_2$ content since a deviation of 60 degrees from a normal incidence upon a
	filament results in a mere factor two difference in the total crossing path.
 
       The question of the origin of these small structures is out of the scope of this paper. 
       Whether their formation is
        triggered by stellar formation processes or whether these clouds are small evaporating pieces of a parent
	body cannot be answered yet. Finally, we note that resilient molecular structures have to be present
	within star forming regions of the MCs and our conclusion follows the one by Dickey et al. (1994) 
	who reported a clear correlation between the cold atomic phase and the star forming regions within
	the LMC.


\subsection{A possible lower limit for the D/H ratio in the SMC of 1.1 $\times$ 10$^{-5}$?}

	Measurements of the atomic D/H ratio have been performed in different astrophysical
	sites, in particular within Damped Lyman Alpha (DLAs) systems observed towards high redshift
	quasar absorbers. The DLAs have generally a few \% solar metallicity and are the best
	candidates to investigate the primordial abundance of deuterium. However, these measurements
	are difficult to achieve and suffer from several systematics, one of which being the
	confusion due to the blend {\bf between} several H\,{\small I} absorbers along crowded lines
	of sight (Lemoine et al. 1999). At this time, definitive
	conclusions are out of reach although a trend towards D/H$_{\rm primordial}$ $\approx$ 3.5 $\times$
	10$^{-5}$ emerges (Kirkman et al. 2001; see for a review Lemoine et al. 1999).
	The D/H ratio has also been thoroughly investigated in the Local ISM (see the review
	by Moos et al. 2002) with the {\small FUSE} satellite. These studies indicate that the deuterium
	abundance is about constant within 100 pc at 1.5 $\times$ 10$^{-5}$ with a standard deviation of
	20 \%. This value is regarded as representative of the current D/H ratio.
	Whether this abundance is representative of the current ISM or whether the LISM has
	peculiar abundances is still a question of debate. At rare occasions, the D/H
	ratio has been measured in the MW along pathlength greater than 1 kpc and was
	consistently found smaller than the LISM value around 0.8 $\times$ 10$^{-5}$
	(Hoopes et al. 2003; H\'ebrard \& Moos 2003).

	The determination of the deuterium abundance in the MCs is of great importance to
	better understand the chemical evolution of the Universe through D/H from the Big Bang Nucleosynthesis (BBN)
	to the current ISM. D/H in the SMC, with two tenths of solar metallicity, would
	allow to have one measurement in the time range between the DLAs formation and the
	current MW ISM. Unfortunately, because the deuterium lines appears
	next to the H\,{\small I} lines, the D\,{\small I} lines are usually severely
	blended by the large galatic H\,{\small I} lines at the SMC velocity.
 
	Another way of investigating the D/H ratio consists in studying the deuterated species present
	in the MC ISM such as DCO$^+$, DCN, ND$_3$ or HD. However, apart from the case of the HD molecule,
	the model used to interpret the fractionation of the deuterium atoms are far from
	being reliable. For instance, recent observations of the ratio NH$_3$/ND$_3$ is in strong
	disagreement with the models, ND$_3$ is found overabundant by a factor 10,000 (van der Tak et al.
	2002). Thus, although one D/H measurement (around 1.5 $\times$ 10$^{-5}$) has been attempted via DCO$^+$
	and DCN by Chin et al. (1996)
	in the SMC, the best probe for the deuterium abundance in the MCs still remains the HD/H$_2$ ratio
	provided there are molecular cores where the deuterium atoms are simply in the form of the HD molecules. 
	In these so--called D--reservoirs HD/H$_2$ $=$ 2 $\times$ D/H.

        At present, from all the  extragalactic {\small FUSE} targets observed to date AV\,95
        is the most promising for that purpose. With a HD/H$_2$ ratio possibly as high as 2.2 $\times$ 10$^{-5}$,
	we can derive D/H of the order of 1.1 $\times$ 10$^{-5}$ that is slightly smaller than the value
	found in the LISM but larger than the value measured towards the two extended MW sightlines
	HD\,191877 and HD\,195695 (Hoopes et al. 2003) with D/H ratios of 0.78 $\times$ 10$^{-5}$ and 0.85 $\times$ 10$^{-5}$
	respectively. For now, unfortunately, the analysis of the HD/H$_2$ ratio towards
	AV\,95 with the existing {\small FUSE} data does not allow us to exclude a much more reduced
	value of HD/H$_2$ of 2 $\times$ 10$^{-6}$. In order to further investigate this interesting sightline
	higher resolution data are mandatory. Such high resolution data could be obtained in the optical
	by a tracer of H$_2$ and HD (Lacour et al. 2004) such as CH at 4300 \AA.
	This observational effort would not only lead to the first clear detection of a deuterium
	reservoir but also help to constrain cosmological models predicting the deuterium
	evolution as a function of metallicity.

\subsection{Molecular filaments detected in the MW}

        LMC and SMC sightlines are especially well suited for the investigation of
        spatial variations in the ISM of our Galaxy (Ferlet 1999). Since the angular distance
between MC background
        stars is much reduced compared with the MW stars, they can be used as
        probe of degree-scale structures in the MW. Two of the stars of our mini-survey
        are close enough to allow for a test of these fine structures: Sk\,$-$68D135 and
Sk\,$-$69D246 are
        within 30$\arcmin$ from each other. The same Na\,{\small I} component is detected
	around $+$25 km s$^{-1}$ with similar $b$-values for both stars.

        At 1 kpc, 30$\arcmin$
        would correspond to a linear distance of 10 pc. We have seen that both clouds have
kinetic temperatures
        around 100 K, as deduced from $T_{01}$, with molecular hydrogen column densities around
$\log$ $N$(H$_2$) $\approx$ 18.
        Then, assuming that these absorbers are within the thick disk which extends at $\pm$ 250
pc from the Galactic plane, the corresponding
        absorber scale is of the order of 2.5 pc or smaller in the direction perpendicular to the
        sightline.
        It is reasonable to assume that we are indeed observing the same molecular structure.

        In the case of Sk\,$-$69D246, the carbon lines detected in the {\small STIS} spectrum give a
	diagnostic of
        the density within the absorber from 60 to 600 atoms cm$^{-3}$ (Table 6).
	Then, from the estimation of the total H$_2$ column density, we
	have a
        rough estimate of the length of the crossed medium in the direction of the sightline which is about 10$^3$ to 10$^4$ AU or 0.1 to 0.01 pc.

        It is clear from these calculations that the molecular structure observed is extremely
filamentary
        with a ratio width/length probably in between 0.4 \% and 4 \%.

        Filamentary structures have been observed for the first time in the Pleiades nebulosity
        (Osterbrock 1959) in optical data and later in radio via the polarization of the H
{\small I}
        lines emitted in the Crab Nebula (Wright 1970). Today, such structures are commonly observed in the
        MW or even in the extragalactic ISM on scales from tens to hundreds of parsecs (Houlahan \& Scalo
1985;
        Kulkarni \& Heiles 1988, Curry 2000). Smaller scales have tentatively been identified
        in H\,{\small I} towards high-velocity pulsars and extragalactic radio sources (Frail et
	al. 1994; Faison et al. 1998) down to the AU scale. Confirmation is brought by recent 
        spectroscopic investigations of the Galactic ISM gas (see Lauroesch, Meyer \& Blades
2000;
        Richter, Sembach \& Howk 2003, Rollinde et al. 2003) although it is not clear yet whether such structures, called
	TSAS for Tiny--Scale Atomic Structures, are filamentary.
        Our present observations add weight, from the observational point of view, to
        the existence and the filamentary nature of the TSMS: Tiny--Scale Molecular Structures.
 
\section{Conclusions}
 
       \begin{enumerate}

	\item We have studied the sightlines towards three LMC bright stars: Sk\,$-$67D05,
	Sk\,$-$68D135, and Sk\,$-$69D246. The combination of ground--based observations
	({\small VLT}) along with archival data from UV satellites ({\small FUSE}, {\small STIS}) makes it possible to investigate
	H$_2$, CO, HD, Cl\,{\small I}, Cl\,{\small II}, C\,{\small I}, C\,{\small I*}, C\,{\small I**}, Na\,{\small I},
	Ca\,{\small I}, Ca\,{\small II}, Fe\,{\small II}, and S\,{\small II}. We also report on the
	investigation of two SMC sightlines with the help of {\small FUSE} and {\small STIS} data. With the exception
	of Na\,{\small I}, Ca\,{\small I}, and Ca\,{\small II}, we have obtained measurements or upper
	limits on all the species observed towards the LMC targets.

	\item In the LMC we find $\langle$HD/H$_2$$\rangle$ $=$ 1.4 ppm,
	\mbox{$\langle$CO/H$_2$ $\rangle$ $=$ 1.4 ppm} (0.8 ppm when excluding the large value obtained toward Sk\,$-$67D05), and
	$\langle$Cl\,{\small I}/H$_2$$\rangle$ $=$ 1.3 ppm. After correcting for the MW reddening we also
	derive $\langle$$N$(H$_2$)/$E_{B-V}$$\rangle$ $=$ 3.5 $\pm$ 1.8 $\times$ 10$^{20}$ mag$^{-1}$ cm$^{-2}$. The analysis of the
	rotational levels of the H$_2$ molecules gives the kinetic temperature of the crossed molecular
	gas about $\langle$$T_{01}$$\rangle$ $\approx$ 70 K. All these values are similar to values obtained
	in the MW disk and seem a bit surprising in the LMC where the radiation field is 5 to 3,000 times
	more intense than in the MW disk and the metallicity down by a factor 2. We suggest that these observations are consistent
	with the existence of cold, dense molecular structures embedded in the diffuse gas.

	\item In the SMC, we find $\langle$$N$(H$_2$)/$E_{B-V}$$\rangle$ $=$ 4.7 and 1.7 $\times$ 10$^{20}$ mag$^{-1}$ cm$^{-2}$
	  for AV\,95 and Sk\,159 respectively. The sightline towards AV\,95 in particular shows unexpectedly large HD/H$_2$
	  (21 ppm) ratio. We tentatively conclude that a D--reservoir is detected
	that can give access to a direct measure of the D/H in the SMC. However, more observations
	are needed, with better S/N and higher resolution, in order to confirm our analysis.

	\item The detailed analysis of the excited states of the carbon atoms gives an interesting
	diagnostic of the H number density along the sightlines towards Sk\,$-$69D246 and AV\,95. For both stars we
	find that the density is greater than 10$^2$ cm$^{-3}$ in the
	LMC molecular components. This leads to clump sizes (or filament) of the order of 0.1 pc. Indeed, we
	note that there are hints for an even denser medium and sizes of the order of 10$^2$
	AU ($\approx$ 0.001 pc). We reached the same conclusions for the Galactic molecular
	component sampled by the Sk\,$-$68D135 and the Sk\,$-$69D246 sightlines. We argue that this
	component probably corresponds to a sheet-like structure such as the ones recently
	inferred in the Galactic halo (Richter et al. 2003).
	More quantitative conclusions are to be given in the following paper based on the chemical
	model of Le Petit et al. (2002). 

	\end{enumerate}

\begin{acknowledgements}
     This work was partially done using the procedure {\it Owens}.f developed
by M. Lemoine and the {\small FUSE} french team. It is our great pleasure to thank 
Eric Maurice for bringing to our attention the PhD thesis of Jocelyne Silvy and
D. Welty for sharing with us some of his {\small VLT} data towards Sk\,159. We gratefully
acknowledge the referee, K.S. de Boer, for his helpful comments.
\end{acknowledgements}

\newpage


   \begin{table*}
      \caption[]{Lines of sight properties and references.}
         \label{Tab1}
     $$ 
         \begin{array}{p{0.2\linewidth}rrlcc}
            \hline
            \noalign{\smallskip}
                        & \textrm{R.A.} & \textrm{Decl.} &  \textrm{Spectral}     &
&         \\
            LMC/SMC stars   &\textrm{J2000} & \textrm{J2000} &   \textrm{Type}        &
\textrm{$E_{(B-V)}$ $^a$} & \textrm{References}\\
            \noalign{\smallskip}
            \hline
            \noalign{\smallskip}
            Sk\,$-$67D05 & 4~50~18.96 & $-$67~39~37.90 & \textrm{O9.7Ib} & 0.16& 1, 2\\
            Sk\,$-$68D135 & 5~37~48.60 & $-$68~55~8.00 & \textrm{ON9.7} Ia+ & 0.27 & 1, 3\\
            Sk\,$-$69D246 & 5~38~53.50 & $-$69~2~0.70  & \textrm{WN6h} & 0.10 & 1, 4\\
            AV\,95      & 0~51~21.54 & $-$72~44~12.90& \textrm{O7 III}& 0.06& 1\\
            Sk\,159     & 1~15~58.84 & $-$73~21~24.10& \textrm{B0.5 Iaw}& 0.09 & 1\\
            \hline
         \end{array}
     $$ 
\begin{list}{}{}
\item[$^{\mathrm{a}}$] Total color excess (MW+LMC)
\item[$^{\mathrm{1}}$] Tumlinson et al. (2002), \item[$^{\mathrm{2}}$] Friedman et al. (2000),
\item[$^{\mathrm{3}}$] Ehrenfreund et al. (2002), \item[$^{\mathrm{4}}$] Bluhm \& DeBoer (2002)
\end{list}
   \end{table*}


\newpage

   \begin{table*}
      \caption[]{Log of {\small FUSE} observations~$^{a}$.}
         \label{Tab2}
     $$ 
         \begin{array}{p{0.2\linewidth}rccc}
            \hline
            \noalign{\smallskip}
            LMC/SMC stars   &  \textrm{Dataset} & \textrm{Date} & \textrm{Exposure time $^b$} &
\textrm{S/N}\\
                        &          &      & \textrm{Ksecs}            & \textrm{per pixel}   \\
            \noalign{\smallskip}
            \hline
            \noalign{\smallskip}
            Sk\,$-$67D05  & P1030703& 2000-10-07 & 3.5 & 5 \\
                      & P1030704& 2000-10-04 & 4.0 & 6 \\
            Sk\,$-$68D135 & P1173901& 2000-02-12 & 6.8 & 5 \\
            Sk\,$-$69D246 & P1031802& 1999-12-16 & 22.1& 9 \\
            AV\,95      & P1150404& 2000-05-31 & 13.8& 7 \\
            Sk\,159     & P1030501& 2000-10-11 & 6.9 & 5 \\
            \hline
         \end{array}
     $$ 
\begin{list}{}{}
\item[$^{\mathrm{a}}$] All these targets have been observed in Time--Tagged (TTAG)mode through the large
aperture (LWRS)
\item[$^{\mathrm{b}}$] After sorting for bad data
\end{list}
   \end{table*}


\newpage

   \begin{table*}
      \caption[]{Log of {\small VLT}/{\small UVES} observations obtained on the 27th May 2003$^*$.}
         \label{Tab3}
     $$ 
         \begin{array}{p{0.08\linewidth}rp{0.1\linewidth}p{0.05\linewidth}p{0.05\linewidth}cl}
            \hline
            \noalign{\smallskip}
             Begin (UT)  &  \textrm{Object} & \textrm{Exposure time (s)} & \textrm{Air Mass Start} & \textrm{Air Mass End} & \textrm{Setup}    \\
            \noalign{\smallskip}
            \hline
            \noalign{\smallskip}
            23:26 & Sk\,$-$69D246 & 1200 & 1.69 & 1.76 & \textrm{Standard Dichroic} \\
            23:50 & Sk\,$-$68D135 & 1200 & 1.78 & 1.86 & \textrm{Standard Dichroic} \\
            00:16 & Sk\,$-$67D05  & 1200 & 2.14 & 2.28 & \textrm{Standard Dichroic} \\
            00:42 & Sk\,$-$68D135 & 489  & 2.02 & 2.06 & \textrm{Non Standard Red} \\
            01:07 & Sk\,$-$69D246 & 900  & 2.16 & 2.27 & \textrm{Non Standard Red} \\
            01:29 & Sk\,$-$67D05  & 513  & 2.79 & 2.89 & \textrm{Non Standard Red} \\
            \hline
         \end{array}
     $$ 
\begin{list}{}{}
\item[$^{\mathrm{*}}$] observations collected at the European Southern Observatory, Paranal, Chile [ESO VLT
69.A-0123(A)].
\end{list}
   \end{table*}



   \begin{table*}
      \caption[]{Log of {\small HST}/{\small STIS} observations.}
         \label{Tab4}
     $$ 
         \begin{array}{p{0.2\linewidth}ccccc}
            \hline
            \noalign{\smallskip}
             Star  &  \textrm{ID} & \textrm{Date} & \textrm{Exposure (s)} & \textrm{Aperture}
\\
                        &          &      &       &         \\
            \noalign{\smallskip}
            \hline
            Sk\,$-$69D246 & \textrm{06LZ12010}& \textrm{2002--11--13}& 1200 &
0\farcs2\times0\farcs2\\
            AV\,95       & \textrm{04WR17010}& \textrm{1999--05--16}& 2448 &
0\farcs2\times0\farcs2\\
                       & \textrm{04WR17020}& \textrm{1999--05--16}& 3168 &
0\farcs2\times0\farcs2 \\
            \hline
         \end{array}
     $$ 
   \end{table*}



\newpage

    \begin{table*}
       \caption[]{Summary of lines used in this work.}
          \label{Tab5}
      $$ 
          \begin{array}{p{0.2\linewidth}lccc}
             \hline
            \noalign{\smallskip}
              \textrm{Species} &  \textrm{Wavelength}  &  \textrm{$f$-value}& \textrm{Dataset}\\
                   &    \textrm{(\AA)}    & &\\
             \noalign{\smallskip}
             \hline
             \noalign{\smallskip}
             \textrm{C\,{\small I}}      & 1276.4822  &  0.449 \times 10^{-2}  &   \textrm{{\small STIS}}\\
             \textrm{C\,{\small I}}      & 1277.2452  &  0.932  \times 10^{-1}  &   \textrm{{\small STIS}}\\
             \textrm{C\,{\small I*}}      & 1276.7498  & 0.252  \times 10^{-2}   &   \textrm{{\small STIS}}\\
             \textrm{C\,{\small I*}}      & 1277.2827  & 0.705 \times  10^{-1}   &   \textrm{{\small STIS}}\\
             \textrm{C\,{\small I*}}      & 1277.5131  & 0.223  \times 10^{-1}   &   \textrm{{\small STIS}}\\
             \textrm{C\,{\small I*}}      & 1279.0562  & 0.736  \times 10^{-3}   &   \textrm{{\small STIS}}\\
             \textrm{C\,{\small I*}}      & 1279.8907  & 0.126  \times 10^{-1}   &   \textrm{{\small STIS}}\\
             \textrm{C\,{\small I**}}      & 1277.1899  & 0.273 \times  10^{-3}   &   \textrm{{\small STIS}}\\
             \textrm{C\,{\small I**}}      & 1277.5500  & 0.791 \times  10^{-1}   &   \textrm{{\small STIS}}\\
             \textrm{C\,{\small I**}}      & 1277.7233  & 0.155 \times  10^{-1}   &   \textrm{{\small STIS}}\\
             \textrm{C\,{\small I**}}      & 1277.9539  & 0.817  \times 10^{-3}   &   \textrm{{\small STIS}}\\
             \textrm{C\,{\small I**}}      & 1279.2290  & 0.378  \times 10^{-2}   &   \textrm{{\small STIS}}\\
             \textrm{C\,{\small I**}}      & 1279.4980  & 0.199  \times 10^{-3}   &   \textrm{{\small STIS}}\\
             \textrm{Ca\,{\small I}}      & 4227.9180  & 0.175 \times 10^{-1}&   \textrm{{\small VLT}}\\
             \textrm{Ca\,{\small II}}      & 3934.7750 & 0.629  \times 10^0&   \textrm{{\small VLT}}\\
             \textrm{Ca\,{\small II}}      & 3969.5901 & 0.312  \times 10^0          &   \textrm{{\small VLT}}\\
             \textrm{Cl\,{\small I}}      & 1088.0590  & 0.810  \times 10^{-1}  &  \textrm{{\small FUSE}}\\
             \textrm{Cl\,{\small I}}      & 1347.2396  & 0.153 \times  10^0&    \textrm{{\small STIS}}\\
             \textrm{Cl\,{\small II}}      & 1071.035  & 0.150  \times 10^{-1}  &  \textrm{{\small FUSE}}\\
             \textrm{Fe\,{\small II}}      & 1125.4478  & 0.160  \times 10^{-1}   &   \textrm{{\small FUSE}}\\
             \textrm{Fe\,{\small II}}      & 1127.0984  & 0.282  \times 10^{-2}   &   \textrm{{\small FUSE}}\\
             \textrm{Fe\,{\small II}}      & 1143.2260  & 0.177  \times 10^{-1}    &  \textrm{{\small FUSE}}\\
             \textrm{Fe\,{\small III}}     & 1122.5240  & 0.544  \times 10^{-1}    &  \textrm{{\small FUSE}}\\
             \textrm{Na\,{\small I}}     & 5891.5833 & 0.641  \times 10^0  &  \textrm{{\small VLT}}\\
             \textrm{Na\,{\small I}}     & 5897.5581 & 0.320  \times 10^0   &   \textrm{{\small VLT}}\\
	     \textrm{HD Lyman (V14--0, {\it J}=0)}	& 959.8174  & 0.147 \times 10^{-1} &  \textrm{{\small FUSE}}\\
	     \textrm{HD Werner (V0--0, {\it J}=0)}	& 1007.2830 & 0.325 \times 10^{-1} &  \textrm{{\small FUSE}}\\
	     \textrm{HD Lyman (V8--0, {\it J}=0)}	& 1011.4570 & 0.262 \times 10^{-1} &  \textrm{{\small FUSE}}\\
	     \textrm{HD Lyman (V5--0, {\it J}=0)}	& 1042.8480 & 0.206 \times 10^{-1} &  \textrm{{\small FUSE}}\\
	     \textrm{HD Lyman (V3--0, {\it J}=0)}	& 1066.2710 & 0.115 \times 10^{-1} &  \textrm{{\small FUSE}}\\
	     \textrm{CO (C--X, {\it J}=0)}		& 1087.8669 & 0.619 \times 10^{-1} &  \textrm{{\small FUSE}}\\
	     \textrm{CO (C--X, {\it J}=1)}		& 1087.9590 & 0.206 \times 10^{-1} &  \textrm{{\small FUSE}}\\
	     \textrm{CO (C--X, {\it J}=1)}		& 1087.8209 & 0.413 \times 10^{-1} &  \textrm{{\small FUSE}}\\
	     \textrm{CO (C--X, {\it J}=2)}		& 1088.0040 & 0.247 \times 10^{-1} &  \textrm{{\small FUSE}}\\
	     \textrm{CO (C--X, {\it J}=2)}		& 1087.7739 & 0.371 \times 10^{-1} &  \textrm{{\small FUSE}}\\
	     \textrm{CO (C--X, {\it J}=3)}		& 1088.0480 & 0.265 \times 10^{-1} &  \textrm{{\small FUSE}}\\
	     \textrm{CO (C--X, {\it J}=3)}		& 1087.7260 & 0.354 \times 10^{-1} &  \textrm{{\small FUSE}}\\
	     \textrm{CO (C--X, {\it J}=4)}		& 1088.0919 & 0.275 \times 10^{-1} &  \textrm{{\small FUSE}}\\
	     \textrm{CO (C--X, {\it J}=4)}		& 1087.6780 & 0.344 \times 10^{-1} &  \textrm{{\small FUSE}}\\
               \hline
             \noalign{\smallskip}
          \end{array}
      $$

   \end{table*}

\newpage

   \begin{table*}
      \caption[]{{\bf Logarithmic column densities} of carbon for two sightlines.}
         \label{Tab6}
     $$ 
         \begin{array}{p{0.1\linewidth}ccc}
            \hline
            \noalign{\smallskip}
                       &\textrm{Sk\,$-$69D246}  & \textrm{AV\,95}\\
                       &                     &    \\
                      & \textrm{Galaxy}& \textrm{Galaxy}\\
            \noalign{\smallskip}
            \hline
            \noalign{\smallskip}
            C\,{\small I}   & 13.40^{+0.08}_{-0.10}& 13.64^{+0.60}_{-0.60}\\
            C\,{\small I}*  & 13.16^{+0.08}_{-0.18}& 13.37^{+0.25}_{-0.25}\\
            C\,{\small I}** & 12.95^{+0.15}_{-0.12}& 12.78^{+0.22}_{-0.18}\\
            $f_1$ $^a$ & 0.30 \pm 0.15 & 0.32 \pm 0.26\\
            $f_2$ $^a$ & 0.18 \pm 0.10 & 0.08 \pm 0.24\\
            $n_{\rm H}$ range $^b$& [60...600]& [30...600]\\
                       &           &     \\
                      &\textrm{\rm LMC}& \textrm{\rm SMC}\\
            \noalign{\smallskip}
            \hline
            \noalign{\smallskip}
            C\,{\small I}   & 13.85^{+0.06}_{-0.06}& 12.90^{+0.27}_{-0.20}\\
            C\,{\small I}*  & 13.72^{+0.10}_{-0.10}& 12.95^{+0.25}_{-0.18}\\
            C\,{\small I}** & 13.44^{+0.05}_{-0.06}& 12.70^{+0.20}_{-0.40}\\
            $f_1$  $^a$& 0.35 \pm 0.07& 0.41 \pm 0.07\\
            $f_2$  $^a$& 0.18 \pm 0.10& 0.23 \pm 0.28\\
            $n_{\rm H}$ range $^c$& [300...600]& [100...800]\\

            \hline
            \noalign{\smallskip}
         \end{array}
     $$ 
\begin{list}{}{}
\item[$^{\mathrm{a}}$] $f_1$ $=$ $N$(C\,{\small I}*)/$N$(C\,{\small I} $_{\rm total}$) and $f_2$ $=$ $N$(C\,{\small
I}**)/$N$(C\,{\small I}$_{\rm total}$).
\item[$^{\mathrm{b}}$] The total density is derived from the model of Jenkins \& Shaya (1979)
assuming the average Galactic radiation field.
\item[$^{\mathrm{c}}$] Same as $^b$ but assuming 10 times the average Galactic radiation field.
\end{list}

   \end{table*}

-----------------------------------------------------------
\newpage

   \begin{table*}
      \caption[]{Ca\,{\small II} and Na\,{\small I} component parameters
      towards the LMC.}
         \label{Tab7}
     $$ 
         \begin{array}{p{0.2\linewidth}lccccccc}
            \hline
            \noalign{\smallskip}
             Star &  \textrm{Comp}  &  \textrm{$V_{\rm hel}$} & \textrm{$N_{10}$} & \textrm{$b$} &
\textrm{$V_{\rm hel}$} & \textrm{$N_{11}$} & \textrm{$b$} \\
                  &        &  \textrm{Ca\,{\small II}}&\textrm{10$^{10}$cm$^{-2}$}&
\textrm{km s$^{-1}$}&\textrm{Na\,{\small I}}
& \textrm{10$^{11}$cm$^{-2}$}& \textrm{km s$^{-1}$}\\
            \noalign{\smallskip}
            \hline
            \noalign{\smallskip}
            Sk\,$-$67D05  & 1 & $-$11 &35.5& 11.2&\\
                      & 2 &0 & 8.6& 3.8 &\\
                      & 3 &6 &31.5& 2.3 &6 & 0.7& 0.4\\
                      & 4 &9 &17.8& 1.7 &\\
                      & 5 &    &    &     &15 & 3.2 & 5.8\\
                      & 6 &18 &56.8& 4.8 &18 & 1.2 & 2.2\\
                      & 7 &    &    &     &22 & 2.6 & 0.8\\
                      & 8 &77 & 4.8& 2.3 &\\
                      & 9 &144 & 6.2& 3.0 &\\
                      & 10&256 & 6.5& 0.7 &\\
                      & 11&266 & 5.3& 1.4 &269 &0.9 & 2.9\\
                      & 12&275 &54.5& 3.0 &274 &3.5 & 1.4\\
                      & 13&284 &62.8& 3.5 &284 &18.0 & 1.6\\
                      & 14&288 &13.9& 3.1 &291 &8.5 & 1.2\\
                      &   &    &    &   \\
            Sk\,$-$68D135 & 1 &15  &71.6& 10.4&19  & 1.7& 0.5\\
                      & 2 &23  &46.2& 1.6 &20  & 3.9& 6.4\\
                      & 3 &    &    &     &27  &12.2& 1.8\\
                      & 4 &63  &146.0&8.5 &\\
                      & 5 &76  &26.2& 2.9 &\\
                      & 6 &    &    &     &84  & 1.2& 12.7\\
                      & 7 &154 &23.0& 4.3 &\\
                      & 8 &166 &31.7& 5.6 &\\
                      & 9 &193 &9.9 & 3.9 &\\
                      & 10&205 &16.0& 7.5 &\\
                      & 11&224 &31.8& 5.4 &\\
                      & 12&241 &27.3& 5.6 &  &    &    &   \\
                      & 13&258 &102.0&8.3 &256 &1.9& 21.1\\
                      & 14&265 &65.1& 2.2 &269 &139.& 0.8\\
                      & 15&272 &160.0&4.9 &277 &75.1& 4.5\\
                      & 16&285 &93.9& 6.6 &289 & 0.8& 1.9\\
                      & 17&303 &20.6& 9.2 &308 & 0.7& 0.2\\
                      &   &    &    &   \\
            Sk\,$-$69D246 & 1 &14  &86.5& 9.9 &14  & 1.9& 2.5\\
                      & 2 &19  &14.6& 1.9 &20  & 3.5& 1.6\\
                      & 3 &    &    &     &25  &34.1& 1.3\\
                      & 4 &34  &17.6& 6.6 &\\
                      & 5 &65  &18.6& 2.9 &67  & 0.4& 0.3\\
                      & 6 &73  &9.2 & 5.3 &\\
                      & 7 &116 &6.1 & 4.4 &\\
                      & 8 &139 & 5.1& 4.3 &\\
                      & 9 &182 &30.4& 17.4&\\
                      & 10&195 & 6.3& 2.3 &\\
                      & 11&218 &24.6& 6.8 &  &    &    &   \\
                      & 12&231 &66.3& 5.2 &  &    &    &   \\
                      & 13&240 &22.3& 2.9 &239 & 0.9& 10.4\\
                      & 14&246 &18.9& 2.6 &248 & 0.2& 0.2\\
                      & 15&254 &40.4& 4.5 &256 & 0.2& 2.0\\
                      & 16&265 &19.8& 5.0 &  &    &    &   \\
                      & 17&276 &42.3& 1.7 &279 &46.1& 3.1\\
                      & 18&280 &217.0&7.8 &285 &17.4& 5.5&   \\
              \hline
            \noalign{\smallskip}
         \end{array}
     $$ 
   \end{table*}


\newpage

   \begin{table*}
      \caption[]{{\bf Logarithmic column densities} of molecular hydrogen derived in the MW.}
         \label{Tab8}
     $$ 
         \begin{array}{p{0.2\linewidth}ccccc}
            \hline
            \noalign{\smallskip}
                       &\textrm{Sk\,$-$67D05}  & \textrm{Sk\,$-$68D135}       & \textrm{Sk\,$-$69D246} &
\textrm{AV\,95}  &\textrm{Sk\,159} \\
            \noalign{\smallskip}
            \hline
            \noalign{\smallskip}
            {\it J}=0 &       14.40^{+0.15}_{-0.30}&  17.84^{+0.11}_{-0.05}&  18.40^{+0.04}_{-0.04}&
18.00^{+0.06}_{-0.06} & 15.70^{+0.30}_{-0.22}\\
            {\it J}=1 &       15.11^{+0.22}_{-0.11}&  18.08^{+0.05}_{-0.04}&  18.46^{+0.04}_{-0.04}&
17.84^{+0.08}_{-0.08}& 16.17^{+0.30}_{-0.10}\\
            {\it J}=2 &       14.30^{+0.24}_{-0.40}&  16.60^{+0.35}_{-0.10}&  16.30^{+0.10}_{-0.07}&
17.00^{+0.18}_{-0.15}& 14.78^{+0.43}_{-0.08}\\
            {\it J}=3 &        <  14.70            &  15.08^{+0.42}_{-0.08}&  15.70^{+0.08}_{-0.08}&
16.00^{+0.08}_{-0.08}& ... \\
            {\it J}=4 &        ...                 &  13.90^{+0.28}_{-0.43}&  14.23^{+0.07}_{-0.06}&
13.90^{+0.30}_{-0.43}& ... \\
             Total &    15.30 \pm 0.17          &       18.29 \pm 0.05          &       18.73
\pm 0.04        &       18.26 \pm 0.05
        &       16.31 \pm 0.25\\
            $T_{01}$ (K) &   136&                 103&
83                 &     66
             &  153\\
            $b_{Dopp}$ (km s$^{-1}$) & 2.9; 2.4 ^a &               4.8 &
4.6         &    1.5      &  2.1
\\
            \hline
            \noalign{\smallskip}
         \end{array}
     $$ 
\begin{list}{}{}
\item[$^{\mathrm{a}}$] Two molecular components are detected towards Sk\,$-$67D05 at $+$35
km s$^{-1}$ and $+$39 km s$^{-1}$. We quote the total H$_2$ column densities.
\end{list}
   \end{table*}



\newpage

   \begin{table*}
      \caption[]{{\bf Logarithmic column densities} of molecular and atomic species detected in the MCs$^a$.}
         \label{Tab9}
     $$ 
         \begin{array}{p{0.2\linewidth}cccccc}
            \hline
            \noalign{\smallskip}
                      &\textrm{Sk\,$-$67D05}   & \textrm{Sk\,$-$68D135}    & \textrm{Sk\,$-$69D246} & \textrm{AV\,95} & \textrm{Sk\,159}\\
            \noalign{\smallskip}
            \hline
            \noalign{\smallskip}
            H$_2$ {\it J}=0 &  19.28^{+0.05}_{-0.06}&  19.46^{+0.10}_{-0.06}&  19.40^{+0.05}_{-0.04} & 19.11^{+0.04}_{-0.04}& 18.85^{+0.06}_{-0.08}\\
            H$_2$ {\it J}=1 & 18.93^{+0.04}_{-0.09}&  19.61^{+0.06}_{-0.04}&  19.30^{+0.04}_{-0.04} & 19.11^{+0.04}_{-0.04}& 18.95^{+0.06}_{-0.08}\\
            H$_2$ {\it J}=2 & 15.60^{+0.30}_{-0.30}&  18.48^{+0.12}_{-0.12}&  17.70^{+0.17}_{-0.50} & 17.81^{+0.05}_{-0.05}& 17.48^{+0.30}_{-0.10}\\
            H$_2$ {\it J}=3 &  15.25^{+0.25}_{-0.15}&  18.23^{+0.07}_{-0.08}&  17.70^{+0.20}_{-0.10} & 17.54^{+0.06}_{-0.07}& 17.03^{+1.30}_{-1.00}\\
            H$_2$ {\it J}=4 & 14.48^{+0.04}_{-0.06}&  17.48^{+0.12}_{-0.05}&  15.49^{+0.22}_{-0.17} & 16.54^{+0.12}_{-0.16}& 13.81^{+1.12}_{-0.40}\\
            H$_2$ {\it J}=5 &  ...    &  16.30^{+0.40}_{-0.07}&  15.17^{+0.45}_{-0.09} & 15.48^{+0.30}_{-0.13}& ... \\
            H$_2$ {\it J}=6 &    ...      &  14.54^{+0.20}_{-0.15}&  13.90^{+0.14}_{-0.12} & ...& ...\\
             Total &    19.44 \pm 0.05  &       19.87 \pm 0.05  &       19.66 \pm 0.04 & 19.43 \pm 0.04 & 19.19 \pm 0.06\\     
            $T_{01}$ (K) &   57&                   92 & 70 & 78 & 78\\
            $b_{Dopp}$ (km s$^{-1}$)&  5.1   &  4.4& 3.6;6.4^b & 0.8 & 1.0;1.3^b\\
            \hline
            \noalign{\smallskip}
         \textrm{$N$(H$_{\rm total}$)}&    21.49   &    21.50               &        21.32        &   21.00       &   ...        \\
            HD {\it J}=0~$^c$ & 13.62^{+0.09}_{-0.12} &14.15^{+0.11}_{-0.15} & 13.62^{+0.07}_{-0.08} & 13.82^{+0.96}_{-0.18}
	    & 13.85^{+0.11}_{-0.14}\\
            CO~$^{\textrm{c}}$ &  13.88^{+0.08}_{-0.09} &13.77^{+0.20}_{-0.28} & 13.57^{+0.08}_{-0.09} &
	    < 13.23 & < 13.34 \\
            Cl\,{\small I} &   13.87^{+0.12}_{-0.13} &13.59^{+0.18}_{-0.20} & 13.46^{+0.24}_{-0.22} & 13.00^{+0.18}_{-0.30}
	    & < 13.32\\
            Cl\,{\small II}& < 13.78       & ...       & 13.73^{+0.14}_{-0.05} & < 14.00 & ...\\
            Fe\,{\small II}&   14.84^{+0.20}_{-0.08} &15.77^{+0.18}_{-0.30} & 15.26^{+0.11}_{-0.35} &  15.22^{+0.50}_{-0.50}
	    & 14.93^{+0.50}_{-0.50}\\
            \hline
            \noalign{\smallskip}
               \textrm{$\chi$ $^d$}&  3         &    3,000               &   20                &   200         &    4         \\
             \textrm{distance (pc)}&     120    &     5                  &   28                &     3         &    50        \\
            \hline
            \noalign{\smallskip}
                 \textrm{[Cl]}&  > 4.64   &    4.09               &     4.60          &   4.00      &     ...      \\
         \textrm{$\delta$(Cl)$^e$}&  > -0.12   &  -0.67             &     -0.16          &   > -0.70     &     ...      \\
\textrm{Cl\,{\small I}/H$_2$(total) (ppm)}&  2.7 \pm 0.9 & 0.5 \pm 0.3 & 0.6 \pm 0.4 & 0.4 \pm 0.3 & < 1.3                     \\
\textrm{Cl\,{\small I}/H$_2$(J=0,1) (ppm)}& 2.7 \pm 0.9 & 0.6 \pm 0.3 & 0.6 \pm 0.4 & 0.4 \pm 0.3 & < 1.3      &              \\
         \textrm{HD/H$_2$(total) (ppm)}&  1.5 \pm 0.5 &  1.9 \pm 0.8  &  0.9 \pm 0.2 & [2.5 ... 22.4] &  4.5 \pm 1.8            \\
         \textrm{CO/H$_2$(total) (ppm)}&  2.7 \pm 0.7 &  0.8 \pm 0.7  &  0.8 \pm 0.2 & < 0.63   &   < 1.33           \\
            \hline
            \noalign{\smallskip}
         \end{array}
     $$ 
\begin{list}{}{}
\item[$^{\mathrm{a}}$] All the errors quoted are 1 $\sigma$. The upper limits are 3 $\sigma$.
\item[$^{\mathrm{b}}$] Two molecular components are detected towards Sk\,$-$69D246 and Sk\,159. 
\item[$^{\mathrm{c}}$] Average between PF and AOD techniques.
\item[$^{\mathrm{d}}$] The Interstellar Radiation Field (ISRF) at 1000 \AA~is given in Draine unit (7 $\times$ 10$^{-17}$ erg cm$^{-3}$ \AA$^{-1}$).
\item[$^{\mathrm{e}}$] The depletion of Chlorine, $\delta$(Cl), is the difference between the LMC and SMC values of 4.76 and 4.70, respectively (see Russell \& Dopita 1992) and the values we derived here.
\end{list}
   \end{table*}


\newpage

   \begin{table*}
      \caption[]{S\,{\small II} and Fe\,{\small II} component parameters toward
        the SMC.}
         \label{Tab10}
     $$ 
         \begin{array}{p{0.2\linewidth}lccccccc}
            \hline
            \noalign{\smallskip}
             Star &  \textrm{Comp}  &  \textrm{$V_{\rm hel}$} & \textrm{$N_{10}$} & \textrm{$b$} &
\textrm{$V_{\rm hel}$} & \textrm{$N_{11}$} & \textrm{$b$} \\
                  &        & \textrm{Fe\,{\small II}}&\textrm{10$^{14}$cm$^{-2}$}&
\textrm{km s$^{-1}$}& \textrm{S\,{\small II}}
& \textrm{10$^{14}$cm$^{-2}$}& \textrm{km s$^{-1}$}\\
            \noalign{\smallskip}
            \hline
            \noalign{\smallskip}
            AV\,95      & 1 &-15&>12.0^a& <15.0^a&     &\\
                      & 2 & 10& 4.7& 8.9 &  7& 11.1& 9.1\\
                      & 3 &    &    &     & 15& 11.3& 5.9\\
                      & 4 & 48&2.4 & 4.7 & 52& 31.0& 0.3 \\
                      & 5 & 75&>0.9^a& <11.^a& 75& 1.0 & 4.4\\
                      & 6 &    &    &     &85&  1.6 & 5.1 \\
                      & 7 & 95& 9.7&17.4 &99&  8.7 & 8.1\\
                      & 8 &123& 7.1&16.8 &119&>20.0^a &<11.0^a\\
                      & 9 &    &    &     &134& 9.0 & 7.7\\
                      & 10&169& 0.9&28.6 &    &     &\\
                      & 11&195&>0.6^a& <21.0^a&\\
                      &   &    &    &   \\
            Sk 159    & 1 &-2.& 1.1&  2.5&\\
                      & 2 & 23& 8.1& 16.9&\\
                      & 3 &150& 8.6& ...^b\\

              \hline
            \noalign{\smallskip}
         \end{array}
     $$ 
\begin{list}{}{}
\item[$^{\mathrm{a}}$] Lower and upper limits quoted are 3 $\sigma$.
\item[$^{\mathrm{b}}$] For this very saturated component the possible $b$-value ranges from 1 to
40 km s$^{-1}$.
\end{list}

   \end{table*}
\clearpage
\newpage
   \begin{figure*}[htbp]
   \centering
   \includegraphics{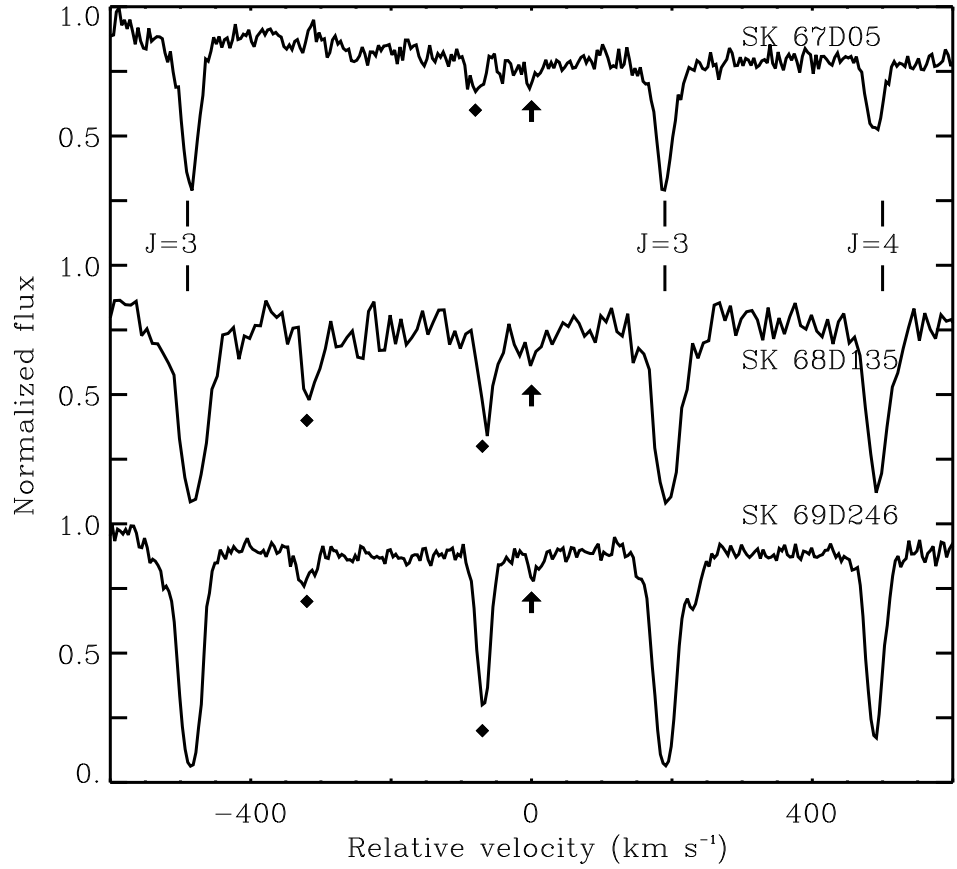}
   \caption{Detection of the 1043 \AA\ HD ($J=$ 0) line in the LMC. H$_2$ Rotational
        levels 3 and 4 are indicated for the LMC component
        while the black diamonds point to the strongest H$_2$ features
        from the Galactic component. The velocities are shifted so that
   the $\lambda$1043 \AA~HD line appears at 0 km s$^{-1}$.}
              \label{Fig1}%
    \end{figure*}

   \begin{figure*}[htbp]
   \centering
   \includegraphics{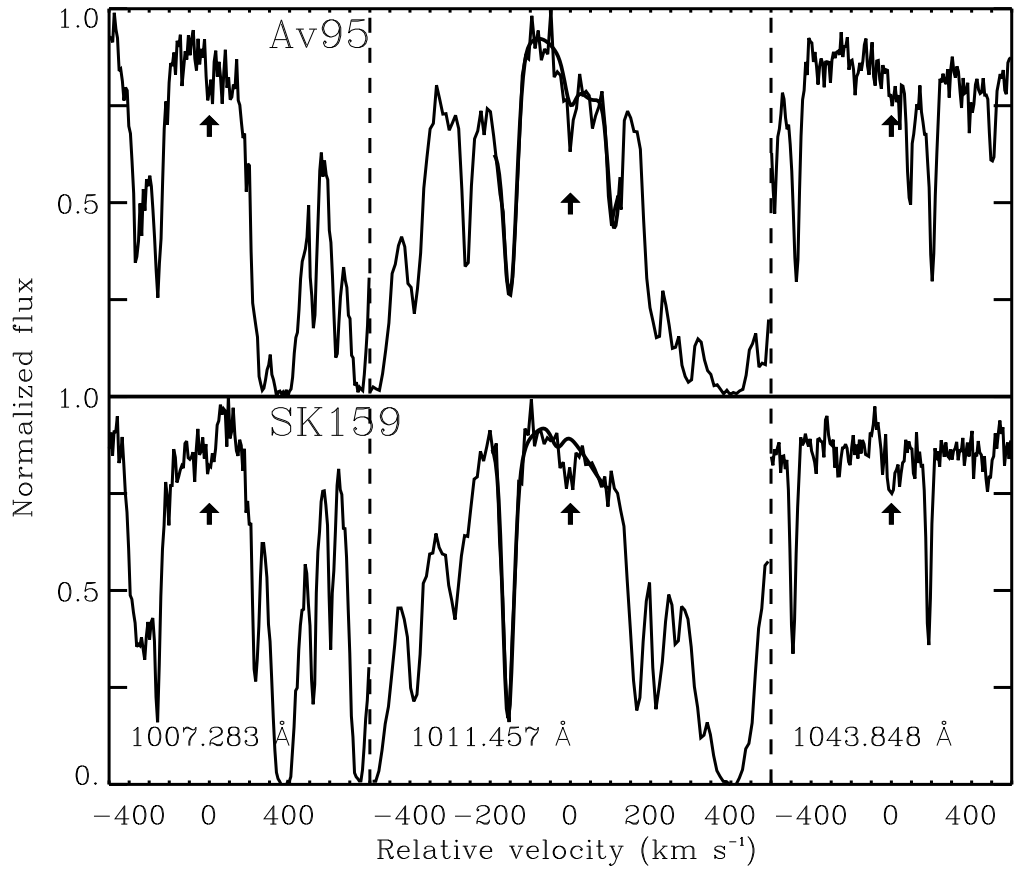}
   \caption{ Detections of the HD molecule towards the SMC at three different wavelengths. AOD analysis
            is based on the 1007 \AA\ HD transition. For clarity we overplotted the
            profile fitting model over the 1011 \AA\ transition which is blended by a Galactic
            H$_2$ line. The velocities are shifted so that the HD lines appear at 0 km s$^{-1}$ in each case.}
              \label{Fig2}%
    \end{figure*}


   \begin{figure*}[htbp]
   \centering
   \includegraphics{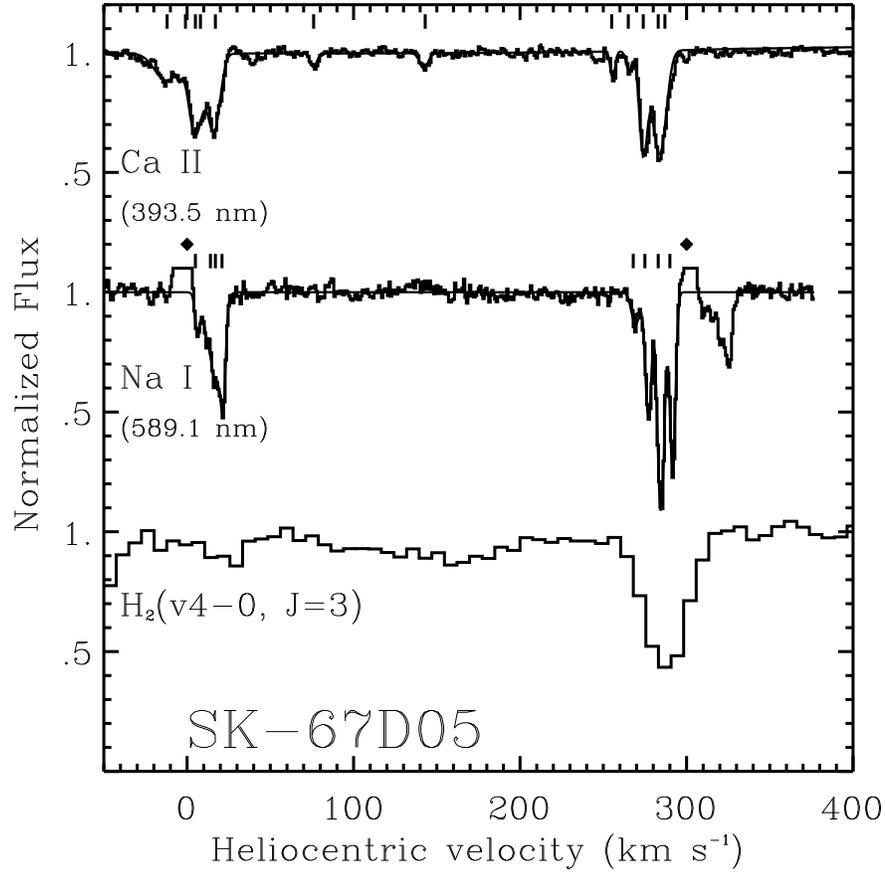}
   \caption{ {\small VLT}/{\small UVES} data for Na and Ca shows up to 14 components, 5 of them arising
           in the MW. Two distincts groups of absorbers show up in the spectra: Galactic at $\approx$
           $+$10 km s$^{-1}$ and LMC at $\approx$ $+$290 km s$^{-1}$. Only two weak intermediate velocity
           absorbers are detected in Ca\,{\small II}. Na {\small I} airglows have been truncated and indicated
           with diamonds. Note that the fits of Na\,{\small I} have been performed on both lines
           of the doublet. Part of the second doublet is seen in the far left of the spectrum displayed.}
              \label{Fig3}%
    \end{figure*}

   \begin{figure*}[htbp]
   \centering
   \includegraphics{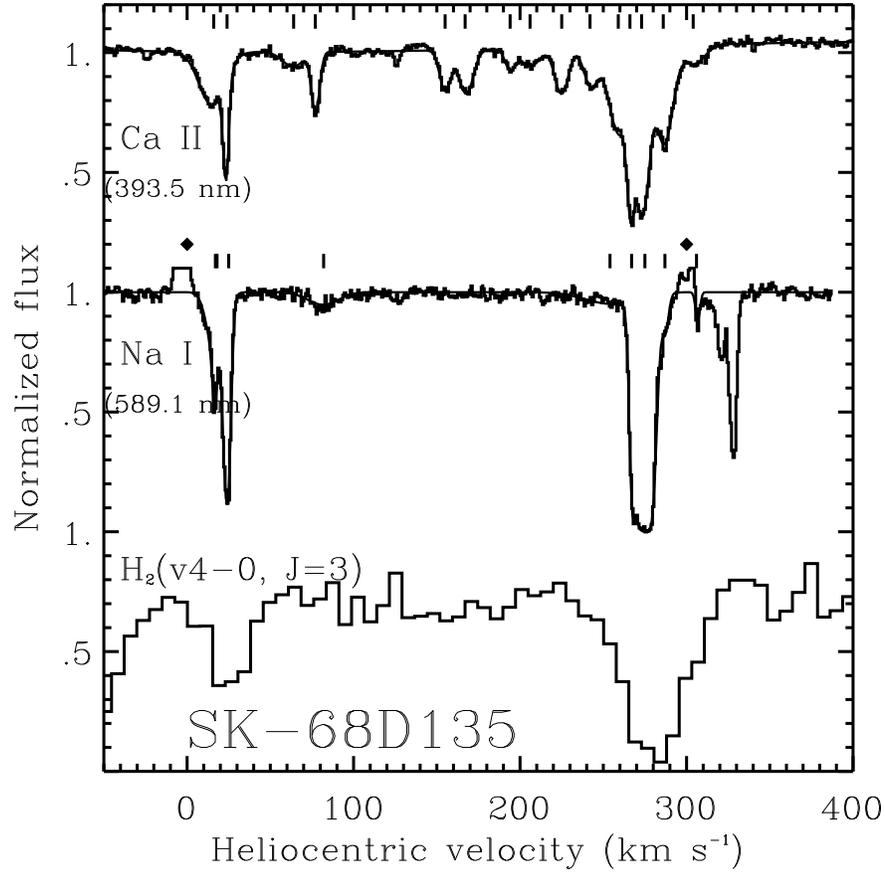}
   \caption{ The sightline towards Sk\,$-$68D135 is similar to the sightline towards Sk\,$-$69D246
            with two groups of strong Na\,{\small I} absorbers: Galaxy and LMC. The {\small FUSE}
            molecular data are consistent with a single molecular absorber in the LMC but
            the resolution is not sufficient to discard the existence of a more complex
	    structure. Note the HVC Na\,{\small I} and Ca\,{\small II} components around $+$310 km s$^{-1}$.
            Na {\small I} airglows have been truncated and indicated with diamonds.}
              \label{Fig4}%
    \end{figure*}

   \begin{figure*}[htbp]
   \centering
   \includegraphics{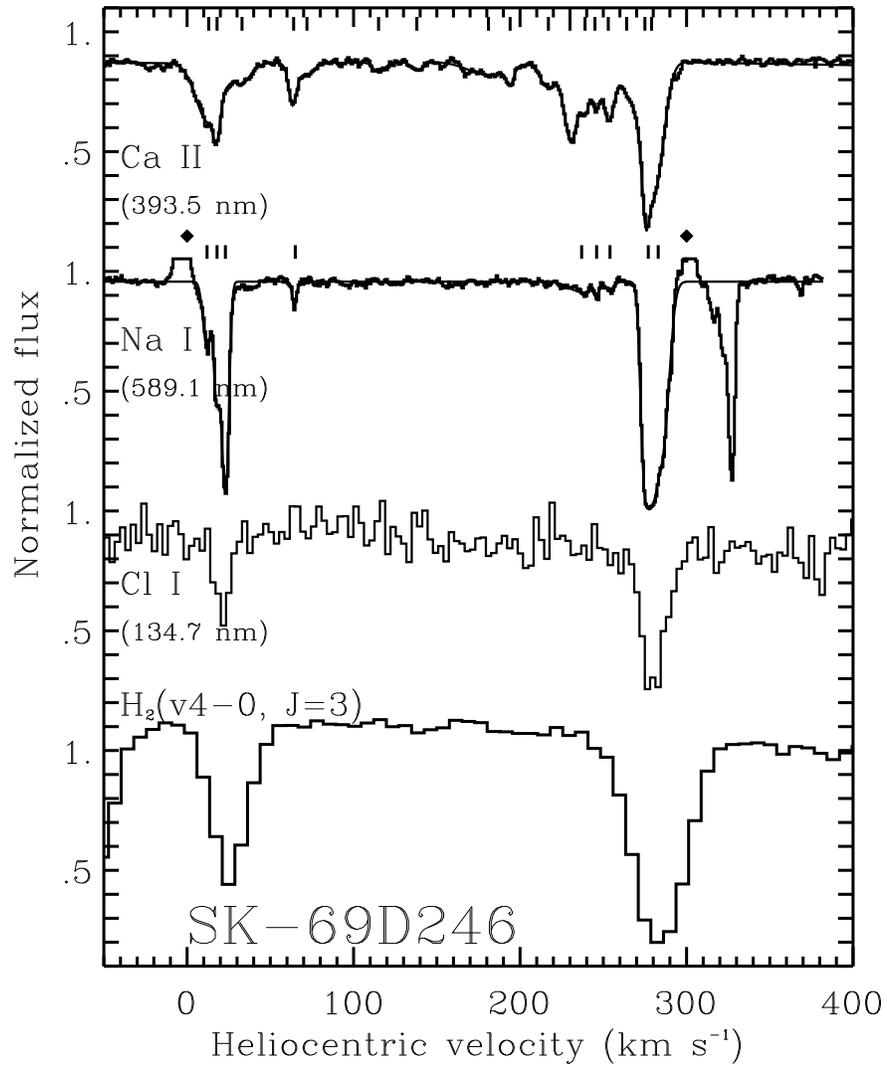}
   \caption{ With 19 components this sightline has the most complicated velocity structure of the
sample.
            Note that with {\small VLT}/{\small UVES}, {\small STIS} and {\small FUSE} data, this sightline is also the one
            for which we have the most complete information about the velocity structure. Note
also that
            the last 2 Ca\,{\small II}, Na\,{\small I} and H$_2$ components coincide more or less
in the
            profile fitting. Na {\small I} airglows have been truncated and indicated with diamonds.}
              \label{Fig5}%
    \end{figure*}


   \begin{figure*}[htbp]
   \centering
   \includegraphics{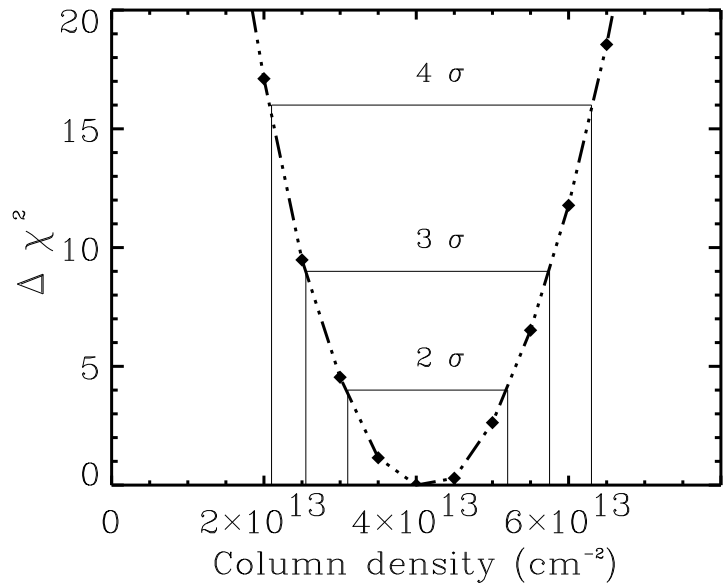}
   \caption{$\chi^2$ plot from the profile fitting of HD towards Sk\,$-$69D246.
                2$\sigma$ error bars are calculated with the standard
                $\Delta \chi^2$ method. Note that for this fit the
                $\overline{\chi^2}$ is very close to unity.}
              \label{Fig6}%
    \end{figure*}


   \begin{figure*}[htbp]
   \centering
   \includegraphics{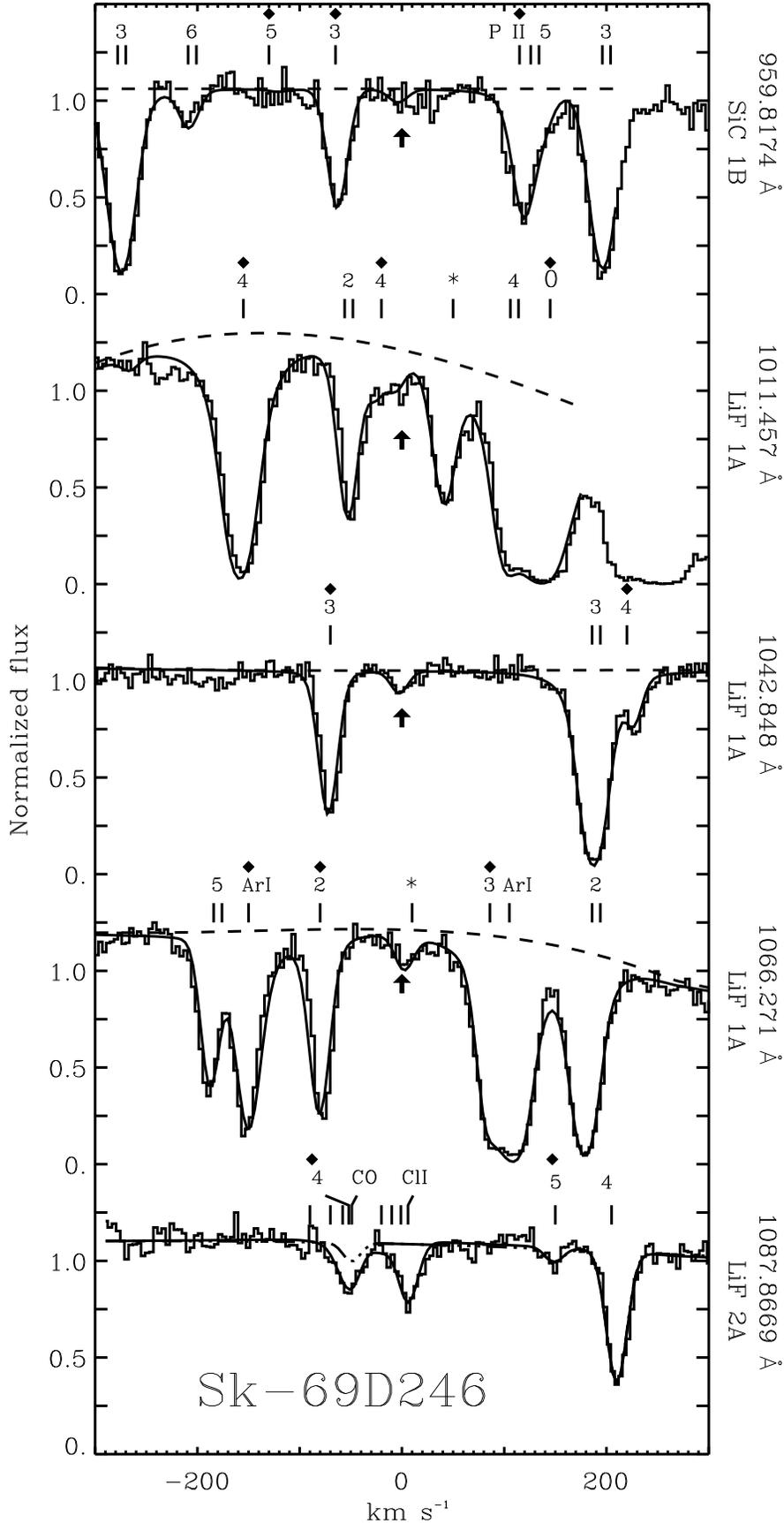}
   \caption{Examples of profiles fitted on the {\small FUSE} data for Sk\,$-$69D246. Galactic absorption lines
        are indicated by the black diamonds and the unidentified stellar features by the stars. 
	The spectra are centered on the LMC HD lines except for the last plot which is centered
	on the LMC Cl\,{\small I} line. Note
        the $J=$ 4 Galactic absorption over the profile of the LMC CO absorption. Note also the
   numerous CO rotational levels (not labeled) spreading over 100 km s$^{-1}$.}
              \label{Fig7}%
    \end{figure*}
   \begin{figure*}[htbp]
   \centering
   \includegraphics{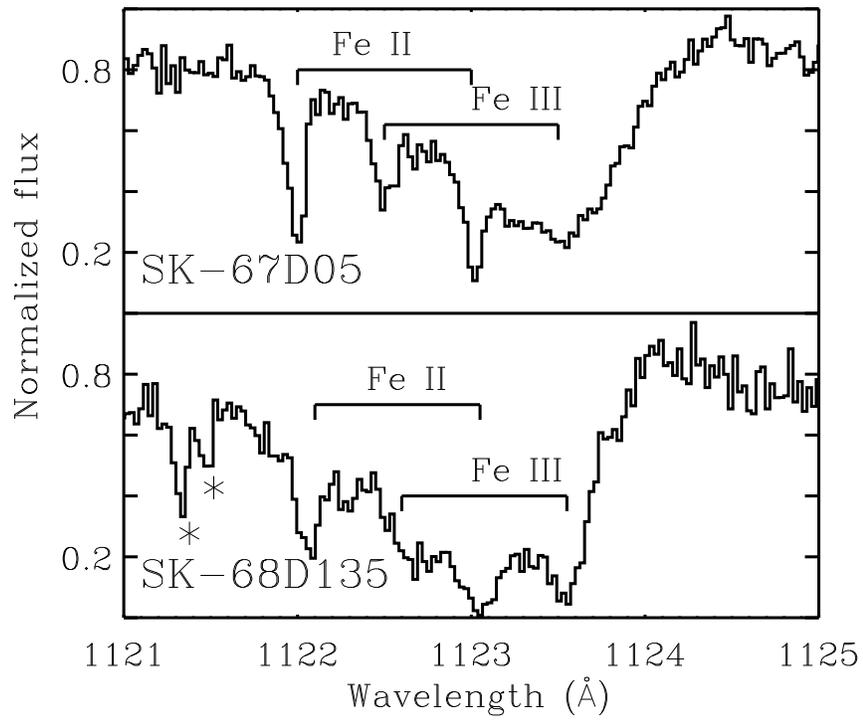}
              \label{FigGam}%
   \caption{ The Fe\,{\small II} and Fe\,{\small III} absorption features arising in
   the MW and in the LMC towards Sk\,$-$68D135 (bottom panel) and Sk\,$-$67D05 (upper panel). Although the LMC
   Fe\,{\small III} line is barely seen towards Sk\,$-$67D05, we can see a strong detection towards Sk\,$-$68D135
   consistent with the intense radiation field.}
              \label{Fig8}%
    \end{figure*}

\newpage
\clearpage
   \begin{figure*}[htbp]
   \centering
   \includegraphics{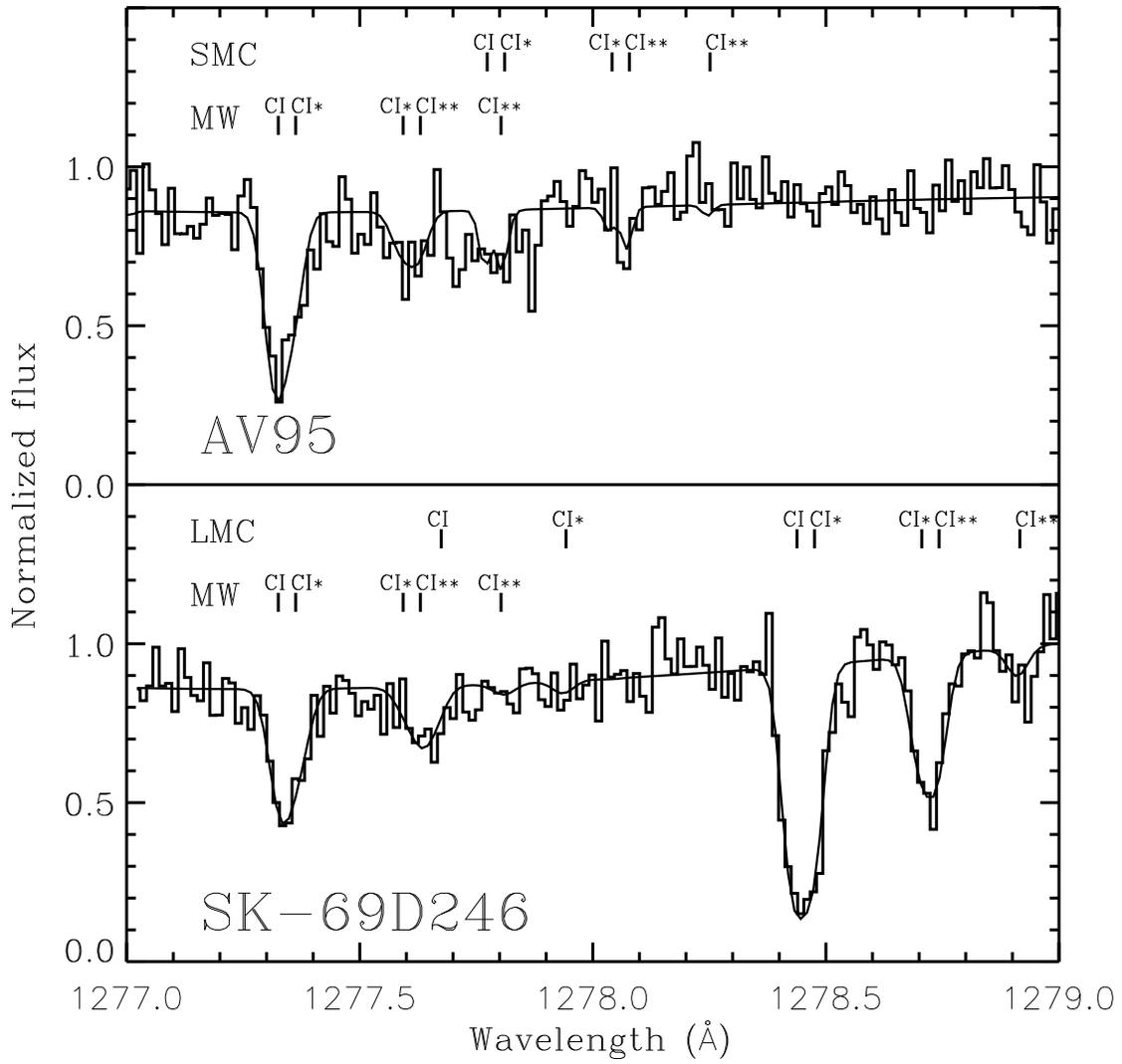}
   \caption{ Profile fitting of the carbon lines detected in the {\small STIS} data towards AV\,95 and
     Sk\,$-$69D246.}
              \label{Fig9}%
    \end{figure*}

\clearpage

   \begin{figure*}[htbp]
   \centering
   \includegraphics{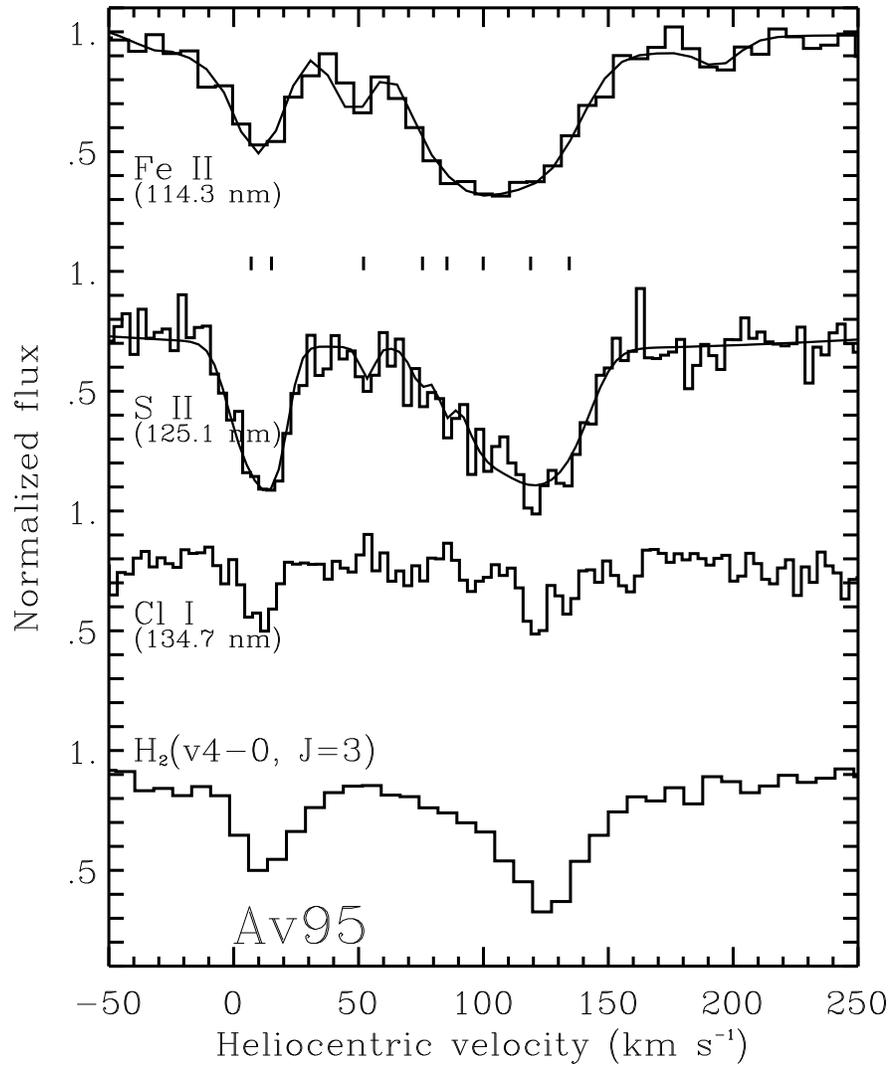}
   \caption{ Atomic and molecular velocity structures towards AV\,95 as derived from {\small STIS} S
{\small II} and
     {\small FUSE} Fe\,{\small II} and H$_2$ data. Up to 11 atomic components are identified
combining sulfur
     and iron absorbers. Note that with a resolution of 20 km s$^{-1}$ many Fe\,{\small II}
absorbers might
     well not be resolved and the Fe\,{\small II} column densities are uncertain. As to the
molecular
     component, observations of Cl\,{\small I} with {\small STIS} at $\approx$ 6 km s$^{-1}$ are
consistent with
     a single molecular cloud.}
              \label{Fig10}%
    \end{figure*}

   \begin{figure*}[htbp]
   \centering
   \includegraphics{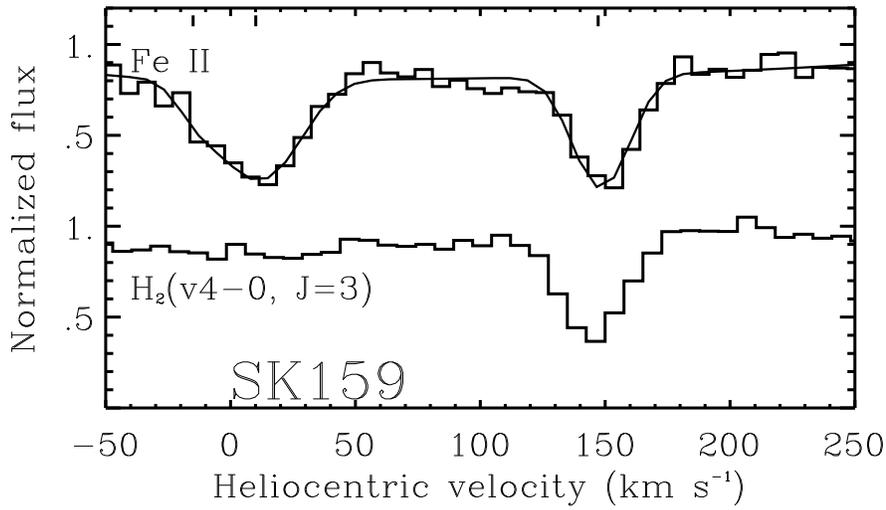}
              \label{FigGam}%
   \caption{ Atomic and molecular velocity structures towards Sk\,159. Based on the {\small FUSE}
data
     itself, only one SMC aborber is revealed. However, higher resolution data, obtained at the
3.6
     CES at ESO Chile, shows that there are at least 2 atomic absorbers present along the sightline at +144 and +149 km s$^{-1}$ (Sylvi 1996).}
              \label{Fig11}%
    \end{figure*}

   \begin{figure*}[htbp]
   \centering
   \includegraphics{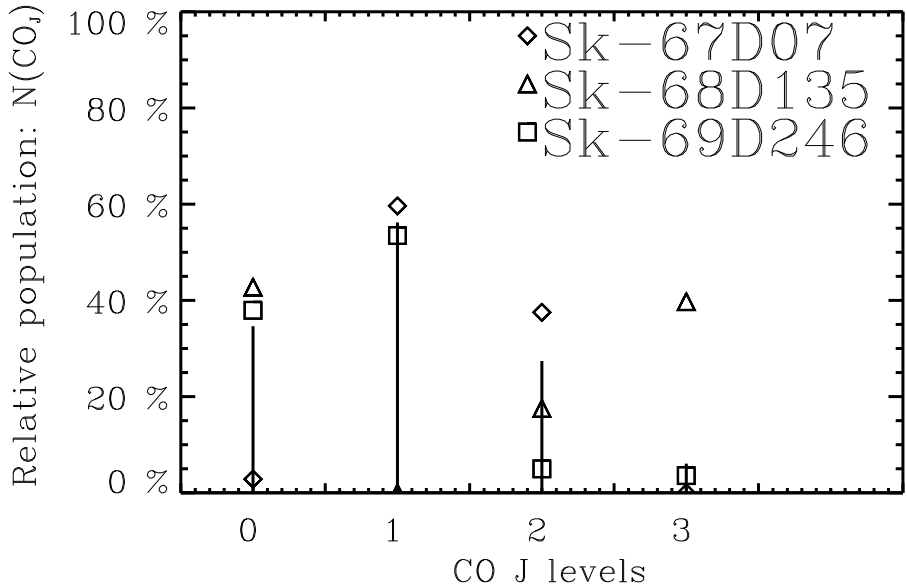}
              \label{FigGam}%
   \caption{ The relative population distribution of the CO $J=$ 1-3 lines derived from the observations is compared
   to a theoretical boltzmann distribution at $T_{ex}$(CO) $=$ 9 K. The agreement is relatively
   good although the error bars on the individual measurements are of the order of 50 \% (not shown).}
              \label{Fig12}%
    \end{figure*}


\begin{thebibliography}{}

	\bibitem[Abgrall et al.(1993)]{1993A&AS..101..273A} Abgrall, H., Roueff, 
	E., Launay, F., Roncin, J.~Y., \& Subtil, J.~L.\ 1993a, \aaps, 101, 273 

	\bibitem[Abgrall et al.(1993)]{1993A&AS..101..323A} Abgrall, H., Roueff, 
	E., Launay, F., Roncin, J.~Y., \& Subtil, J.~L.\ 1993b, \aaps, 101, 323 
	
	\bibitem[Abgrall, Roueff, \& Drira(2000)]{2000A&AS..141..297A} Abgrall, H., 
	Roueff, E., \& Drira, I.\ 2000, \aaps, 141, 297 

	\bibitem[Abgrall \& Roueff(2003)]{2000A&AS..141..prep} Abgrall, H., 
	Roueff, E. 2003 in preparation

	\bibitem[Ballester et al.(2000)]{2000SPIE.4010..246B} Ballester, P., 
	Grosbol, P., Banse, K., Disaro, A., Dorigo, D., Modigliani, A., Pizarro de 
	la Iglesia, J.~A., \& Boitquin, O.\ 2000, \procspie, 4010, 246 

	\bibitem[Bertin, Lallement, Ferlet, \& 
	Vidal-Madjar(1993)]{1993A&A...278..549B} Bertin, P., Lallement, R., Ferlet, 
	R., \& Vidal-Madjar, A.\ 1993, \aap, 278, 549 

        \bibitem[Bluhm \& de Boer(2001)]{2001A&A...379...82B} Bluhm, H.~\& de Boer, 
        K.~S.\ 2001, \aap, 379, 82

        \bibitem[Boehringer \& Hartquist(1987)]{1987MNRAS.228..915B} Boehringer, 
        H.~\& Hartquist, T.~W.\ 1987, \mnras, 228, 915 

        \bibitem[Bolatto, Jackson, \& Ingalls(1999)]{1999ApJ...513..275B} Bolatto, 
	  A.~D., Jackson, J.~M., \& Ingalls, J.~G.\ 1999, \apj, 513, 275 

       \bibitem[Cardelli, Clayton, \& Mathis(1989)]{1989ApJ...345..245C} Cardelli, 
        J.~A., Clayton, G.~C., \& Mathis, J.~S.\ 1989, \apj, 345, 245 

      \bibitem[Cecchi-Pestellini \& Dalgarno(2002)]{2002MNRAS.331L..31C} 
	Cecchi-Pestellini, C.~\& Dalgarno, A.\ 2002, \mnras, 331, L31 

      \bibitem[Cecchi-Pestellini \& Dalgarno(2000)]{2000MNRAS.313L...6C} 
	Cecchi-Pestellini, C.~\& Dalgarno, A.\ 2000, \mnras, 313, L6 

	\bibitem[Chin et al.(1996)]{1996A&A...312L..33C} Chin, Y.-N., Henkel, C., 
	Millar, T.~J., Whiteoak, J.~B., \& Mauersberger, R.\ 1996, \aap, 312, L33 

	\bibitem[Chu, Wakker, Mac Low, \& Garcia-Segura(1994)]{1994AJ....108.1696C} 
	Chu, Y.-H., Wakker, B., Mac Low, M.-M., \& Garcia-Segura, G.\ 1994, \aj, 
	108, 1696 

        \bibitem[Cohen et al.(1988)]{1988ApJ...331L..95C} Cohen, R.~S., Dame, 
	  T.~M., Garay, G., Montani, J., Rubio, M., \& Thaddeus, P.\ 1988, \apjl, 
	  331, L95 

	\bibitem[Curry(2000)]{2000ApJ...541..831C} Curry, C.~L.\ 2000, \apj, 541, 
	831 

	\bibitem[Curry(2002)]{2002ApJ...576..849C} Curry, C.~L.\ 2002, \apj, 576, 
	849 

	\bibitem[Danforth et al.(2002)]{2002ApJS..139...81D} Danforth, C.~W., Howk, 
	J.~C., Fullerton, A.~W., Blair, W.~P., \& Sembach, K.~R.\ 2002, \apjs, 139, 
	81 

        \bibitem[de Boer, Koornneef, \& Savage(1980)]{1980ApJ...236..769D} de Boer, 
        K.~S., Koornneef, J., \& Savage, B.~D.\ 1980, \apj, 236, 769 

	\bibitem[Dekker et al.(2000)]{2000SPIE.4008..534D} Dekker, H., D'Odorico, 
	S., Kaufer, A., Delabre, B., \& Kotzlowski, H.\ 2000, \procspie, 4008, 534 

      \bibitem[Dickey et al.(1994)]{1994A&A...289..357D} Dickey, J.~M., Mebold, 
	U., Marx, M., Amy, S., Haynes, R.~F., \& Wilson, W.\ 1994, \aap, 289, 357 

      \bibitem[Diplas \& Savage(1994)]{1994ApJ...427..274D} Diplas, A.~\& Savage, 
	B.~D.\ 1994, \apj, 427, 274 

	\bibitem[de Vaucouleurs \& Freeman(1972)]{1972VA.....14..163D} de 
	Vaucouleurs, G.~\& Freeman, K.~C.\ 1972, Vistas in Astronomy, 14, 163 	B.~D.\ 1994, \apj, 427, 274 

        \bibitem[Draine(1978)]{1978ApJS...36..595D} Draine, B.~T.\ 1978, \apjs, 36, 
	  595 

	\bibitem[Dufour, Shields, \& Talbot(1982)]{1982ApJ...252..461D} Dufour, 
	R.~J., Shields, G.~A., \& Talbot, R.~J.\ 1982, \apj, 252, 461 

        \bibitem[Ehrenfreund et al.(2002)]{2002ApJ...576L.117E} Ehrenfreund, P.~et 
        al.\ 2002, \apjl, 576, L117 

	\bibitem[Faison, Goss, Diamond, \& Taylor(1998)]{1998AJ....116.2916F} 
	Faison, M.~D., Goss, W.~M., Diamond, P.~J., \& Taylor, G.~B.\ 1998, \aj, 
	116, 2916 

	\bibitem[Ferlet, Vidal-Madjar, \& Gry(1985)]{1985ApJ...298..838F} Ferlet, 
	R., Vidal-Madjar, A., \& Gry, C.\ 1985, \apj, 298, 838 

	\bibitem[Ferlet(1999)]{1999A&ARv...9..153F} Ferlet, R.\ 1999, \aapr, 9, 153 
        
	\bibitem[Ferlet et al.(2000)]{2000ApJ...538L..69F} Ferlet, R.~et al.\ 2000, 
        \apjl, 538, L69

	\bibitem[Fitzpatrick \& Savage(1984)]{1984ApJ...279..578F} Fitzpatrick, 
	E.~L.~\& Savage, B.~D.\ 1984, \apj, 279, 578 

	\bibitem[Frail, Weisberg, Cordes, \& Mathers(1994)]{1994ApJ...436..144F} 
	Frail, D.~A., Weisberg, J.~M., Cordes, J.~M., \& Mathers, C.\ 1994, \apj, 
	436, 144 

        \bibitem[Friedman et al.(2000)]{2000ApJ...538L..39F} Friedman, S.~D.~et 
        al.\ 2000, \apjl, 538, L39

	\bibitem[Garay et al.(2002)]{2002A&A...389..977G} Garay, G., Johansson, 
	L.~E.~B., Nyman, L.-{\AA}., Booth, R.~S., Israel, F.~P., Kutner, M.~L., 
	Lequeux, J., \& Rubio, M.\ 2002, \aap, 389, 977 

	\bibitem[Garnett et al.(1999)]{1999ApJ...513..168G} Garnett, D.~R., 
	Shields, G.~A., Peimbert, M., Torres-Peimbert, S., Skillman, E.~D., Dufour, 
	R.~J., Terlevich, E., \& Terlevich, R.~J.\ 1999, \apj, 513, 168 

        \bibitem[H{\' e}brard et al.(2002)]{2002ApJS..140..103H} H{\' e}brard, 
        G.~et al.\ 2002, \apjs, 140, 103 
        
      \bibitem[H{\' e}brard \& Moos(2003)]{2003ApJ...599..297H} H{\' e}brard, 
	G.~\& Moos, H.~W.\ 2003, \apj, 599, 297 

        \bibitem[Herbig(1995)]{1995ARA&A..33...19H} Herbig, G.~H.\ 1995, \araa, 33, 
        19 

	\bibitem[Hoopes et al.(2003)]{2003ApJ...586.1094H} Hoopes, C.~G., Sembach, 
	K.~R., H{\' e}brard, G., Moos, H.~W., \& Knauth, D.~C.\ 2003, \apj, 586, 
	1094 

	\bibitem[Houlahan \& Scalo(1985)]{1985BAAS...17R.835H} Houlahan, P.~\& 
	Scalo, J.~M.\ 1985, \baas, 17, 835 

	\bibitem[Howk, Savage, Sembach, \& Hoopes(2002)]{2002ApJ...572..264H} Howk, 
	J.~C., Savage, B.~D., Sembach, K.~R., \& Hoopes, C.~G.\ 2002, \apj, 572, 
	264 

      \bibitem[Israel(1991)]{1991IAUS..146...13I} Israel, F.~P.\ 1991, IAU 
	Symp.~146: Dynamics of Galaxies and Their Molecular Cloud Distributions, 
	146, 13 

	\bibitem[Israel et al.(2003)]{2003A&A...401...99I} Israel, F.~P.~et al.\ 
	2003, \aap, 401, 99 

	\bibitem[Jenkins \& Shaya(1979)]{1979ApJ...231...55J} Jenkins, E.~B.~\& 
	Shaya, E.~J.\ 1979, \apj, 231, 55 

	\bibitem[Jenkins, Savage, \& Spitzer(1986)]{1986ApJ...301..355J} Jenkins, 
	E.~B., Savage, B.~D., \& Spitzer, L.\ 1986, \apj, 301, 355 

	\bibitem[Jenkins \& Tripp(2001)]{2001ApJS..137..297J} Jenkins, E.~B.~\& 
	Tripp, T.~M.\ 2001, \apjs, 137, 297 

        \bibitem[Jura(1974)]{1974ApJ...190L..33J} Jura, M.\ 1974, \apjl, 190, L33 

        \bibitem[Jura \& York(1978)]{1978ApJ...219..861J} Jura, M.~\& York, D.~G.\ 
        1978, \apj, 219, 861 

	\bibitem[Kimble et al.(1998)]{1998SPIE.3356..188K} Kimble, R.~A.~et al.\ 
	1998, \procspie, 3356, 188 

	\bibitem[Kirkman et al.(2001)]{2001ApJ...559...23K} Kirkman, D.~et al.\ 
	2001, \apj, 559, 23 

	\bibitem[Kobulnicky \& Dickey(1999)]{1999AJ....117..908K} Kobulnicky, 
	H.~A.~\& Dickey, J.~M.\ 1999, \aj, 117, 908 

	\bibitem[Koornneef(1982)]{1982A&A...107..247K} Koornneef, J.\ 1982, \aap, 
	107, 247 

	\bibitem[Kulkarni \& Heiles(1988)]{1988gera.book...95K} Kulkarni, S.~R.~\& 
	Heiles, C.\ 1988, Galactic and Extragalactic Radio Astronomy, 95 

	\bibitem[Lauroesch, Meyer, \& Blades(2000)]{2000ApJ...543L..43L} Lauroesch, 
	J.~T., Meyer, D.~M., \& Blades, J.~C.\ 2000, \apjl, 543, L43 

	\bibitem[Lacour et al. (2004)]{2000ApJ...543L..43L} Lacour, S., Le Petit, 
	F., Sonnentrucker, P., Andr{\' e}, M., D{\' e}sert, J.~M., Ferlet, R., \& 
	Roueff, 2004 in preparation

	\bibitem[Lehner(2002)]{2002ApJ...578..126L} Lehner, N.\ 2002, \apj, 578, 
	126 

        \bibitem[Lehner et al.(2002)]{2002ApJS..140...81L} Lehner, N., Gry, C., 
	  Sembach, K.~R., H{\' e}brard, G., Chayer, P., Moos, H.~W., Howk, J.~C., \& 
	  D{\' e}sert, J.-M.\ 2002, \apjs, 140, 81 

	\bibitem[Lemoine et al.(1999)]{1999NewA....4..231L} Lemoine, M.~et al.\ 
	1999, New Astronomy, 4, 231 

        \bibitem[Lemoine et al.(2002)]{2002ApJS..140...67L} Lemoine, M.~et al.\ 
        2002, \apjs, 140, 67 

        \bibitem[Le Petit, Roueff, \& Le Bourlot(2002)]{2002A&A...390..369L} Le 
        Petit, F., Roueff, E., \& Le Bourlot, J.\ 2002, \aap, 390, 369 


        \bibitem[Lequeux et al.(1994)]{1994A&A...292..371L} Lequeux, J., Le 
        Bourlot, J., Des Forets, G.~P., Roueff, E., Boulanger, F., \& Rubio, M.\ 
        1994, \aap, 292, 371 

	\bibitem[Mallouris et al.(2001)]{2001ApJ...558..133M} Mallouris, C.~et al.\ 
	2001, \apj, 558, 133 

        \bibitem[McKee \& Ostriker(1977)]{1977ApJ...218..148M} McKee, C.~F.~\& 
        Ostriker, J.~P.\ 1977, \apj, 218, 148 
        
      \bibitem[Mebold et al.(1991)]{1991A&A...251L...1M} Mebold, U., Herbstmeier, 
	U., Kalberla, P.~M.~W., Greisen, E.~W., Wilson, W., \& Haynes, R.~F.\ 1991, 
	\aap, 251, L1 

        \bibitem[Moos et al.(2000)]{2000ApJ...538L...1M} Moos, H.~W.~et al.\ 2000, 
        \apjl, 538, L1

        \bibitem[Morton \& Noreau(1994)]{1994ApJS...95..301M} Morton, D.~C.~\& 
	  Noreau, L.\ 1994, \apjs, 95, 301 

	\bibitem[Neubig \& Bruhweiler(1997)]{1997AJ....114.1951N} Neubig, 
	M.~M.~S.~\& Bruhweiler, F.~C.\ 1997, \aj, 114, 1951 

	\bibitem[Osterbrock(1959)]{1959PASP...71...23O} Osterbrock, D.~E.\ 1959, 
	\pasp, 71, 23 

        \bibitem[Parravano, Hollenbach, \& McKee(2003)]{2003ApJ...584..797P} 
	  Parravano, A., Hollenbach, D.~J., \& McKee, C.~F.\ 2003, \apj, 584, 797 


	\bibitem[P\'equignot \& Aldrovandi(1986)]{1986A&A...161..169P} P\'equignot, 
	   D.~\& Aldrovandi, S.~M.~V.\ 1986, \aap, 161, 169 

        \bibitem[Pfenniger, Combes, \& Martinet(1994)]{1994A&A...285...79P} 
        Pfenniger, D., Combes, F., \& Martinet, L.\ 1994, \aap, 285, 79 

	\bibitem[Pottasch(1972)]{1972A&A....20..245P} Pottasch, S.~R.\ 1972, \aap, 
	20, 245 
        
        \bibitem[Rachford et al.(2001)]{2001ApJ...555..839R} Rachford, B.~L.~et 
	al.\ 2001, \apj, 555, 839 

	\bibitem[Rachford et al.(2002)]{2002ApJ...577..221R} Rachford, B.~L.~et 
        al.\ 2002, \apj, 577, 221 
	
	\bibitem[Richter et al.(1999)]{1999Natur.402..386R} Richter, P., de Boer, 
	K.~S., Widmann, H., Kappelmann, N., Gringel, W., Grewing, M., \& Barnstedt, 
	J.\ 1999, \nat, 402, 386 

       \bibitem[Richter, Sembach, \& Howk(2003)]{2003A&A...405.1013R} Richter, P., 
	Sembach, K.~R., \& Howk, J.~C.\ 2003, \aap, 405, 1013 


      \bibitem[Rollinde, Boiss{\' e}, Federman, \& 
	Pan(2003)]{2003A&A...401..215R} Rollinde, E., Boiss{\' e}, P., Federman, 
	S.~R., \& Pan, K.\ 2003, \aap, 401, 215 

       \bibitem[Roueff \& Zeippen(2000)]{2000A&AS..142..475R} Roueff, E.~\& 
	Zeippen, C.~J.\ 2000, \aaps, 142, 475 

	\bibitem[Russell \& Dopita(1992)]{1992ApJ...384..508R} Russell, S.~C.~\& 
	Dopita, M.~A.\ 1992, \apj, 384, 508 

      \bibitem[Sahnow et al.(2000)]{2000ApJ...538L...7S} Sahnow, D.~J.~et al.\ 
	2000, \apjl, 538, L7 

      \bibitem[Savage \& de Boer(1981)]{1981ApJ...243..460S} Savage, B.~D.~\& de 
	Boer, K.~S.\ 1981, \apj, 243, 460 

	\bibitem[Savage \& Sembach(1991)]{1991ApJ...379..245S} Savage, B.~D.~\& 
	  Sembach, K.~R.\ 1991, \apj, 379, 245 

       \bibitem[Savage, Apponi, Ziurys, \& Wyckoff(2002)]{2002ApJ...578..211S} 
	  Savage, C., Apponi, A.~J., Ziurys, L.~M., \& Wyckoff, S.\ 2002, \apj, 578, 
	  211 

	\bibitem[Schectman, Federman, Beideck, \& Ellis(1993)]{1993ApJ...406..735S} 
	Schectman, R.~M., Federman, S.~R., Beideck, D.~J., \& Ellis, D.~J.\ 1993, 
	\apj, 406, 735 

	\bibitem[Schramm \& Turner(1998)]{1998RvMP...70..303S} Schramm, D.~N.~\& 
	Turner, M.~S.\ 1998, Reviews of Modern Physics, 70, 303 

       \bibitem[Sembach \& Savage(1992)]{1992ApJS...83..147S} Sembach, K.~R.~\& 
	 Savage, B.~D.\ 1992, \apjs, 83, 147 

      \bibitem[Schectman, Federman, Beideck, \& Ellis(1993)]{1993ApJ.} Silvy, J. 1996, PhD Universit\'e de Marseille

      \bibitem[Snow \& Cohen(1974)]{1974ApJ...194..313S} Snow, T.~P.~\& Cohen, 
        J.~G.\ 1974, \apj, 194, 313
 
	\bibitem[Snowden \& Petre(1994)]{1994ApJ...436L.123S} Snowden, S.~L.~\& 
	Petre, R.\ 1994, \apjl, 436, L123
	
        \bibitem[Songaila(1981)]{1981ApJ...248..945S} Songaila, A.\ 1981, \apj, 
        248, 945 

        \bibitem[Sonneborn et al.(2002)]{2002ApJS..140...51S} Sonneborn, G.~et al.\ 
        2002, \apjs, 140, 51 

        \bibitem[Sonnentrucker et al.(2002)]{2002ApJ...576..241S} Sonnentrucker, 
        P., Friedman, S.~D., Welty, D.~E., York, D.~G., \& Snow, T.~P.\ 2002, \apj, 
        576, 241 

        \bibitem[Sonnentrucker et al.(2003)]{2003ApJ...596..350S} Sonnentrucker, 
	  P., Friedman, S.~D., Welty, D.~E., York, D.~G., \& Snow, T.~P.\ 2003, \apj, 
	  596, 350 

        \bibitem[Spitzer \& Cochran(1973)]{1973ApJ...186L..23S} Spitzer, L.~J.~\& 
        Cochran, W.~D.\ 1973, \apjl, 186, L23 

        \bibitem[Spitzer \& Zweibel(1974)]{1974ApJ...191L.127S} Spitzer, L.~J.~\& 
        Zweibel, E.~G.\ 1974, \apjl, 191, L127 

        \bibitem[Tumlinson et al.(2002)]{2002ApJ...566..857T} Tumlinson, J.~et al.\ 
        2002, \apj, 566, 857 

	\bibitem[van der Tak et al.(2002)]{2002A&A...388L..53V} van der Tak, 
	F.~F.~S., Schilke, P., M{\" u}ller, H.~S.~P., Lis, D.~C., Phillips, T.~G., 
	Gerin, M., \& Roueff, E.\ 2002, \aap, 388, L53 

	\bibitem[Varshalovich et al.(2001)]{2001AstL...27..683V} Varshalovich, 
	D.~A., Ivanchik, A.~V., Petitjean, P., Srianand, R., \& Ledoux, C.\ 2001, 
	Astronomy Letters, 27, 683 

	\bibitem[Vidal-Madjar et al.(1987)]{1987A&A...177L..17V} Vidal-Madjar, A., 
	Andreani, P., Cristiani, S., Ferlet, R., Lanz, T., \& Vladilo, G.\ 1987, 
	\aap, 177, L17 

        \bibitem[Vladilo et al.(1993)]{1993A&A...274...37V} Vladilo, G., Molaro, 
        P., Monai, S., D'Odorico, S., Ferlet, R., Vidal--Madjar, A.~V., \& Dennefeld, M.\ 
        1993, \aap, 274, 37 

        \bibitem[Wayte(1990)]{1990ApJ...355..473W} Wayte, S.~R.\ 1990, \apj, 355, 
        473     

	\bibitem[Welty, Morton, \& Hobbs(1996)]{1996ApJS..106..533W} Welty, D.~E., 
	Morton, D.~C., \& Hobbs, L.~M.\ 1996, \apjs, 106, 533 

	\bibitem[Welty et al.(1997)]{1997ApJ...489..672W} Welty, D.~E., Lauroesch, 
	J.~T., Blades, J.~C., Hobbs, L.~M., \& York, D.~G.\ 1997, \apj, 489, 672 

	\bibitem[Welty, Frisch, Sonneborn, \& York(1999)]{1999ApJ...512..636W} 
	Welty, D.~E., Frisch, P.~C., Sonneborn, G., \& York, D.~G.\ 1999, \apj, 
	512, 636 

	\bibitem[Welty \& Hobbs(2001)]{2001ApJS..133..345W} Welty, D.~E.~\& Hobbs, 
	L.~M.\ 2001, \apjs, 133, 345 

        \bibitem[Wood et al.(2002)]{2002ApJS..140...91W} Wood, B.~E., Linsky, 
        J.~L., H{\' e}brard, G., Vidal-Madjar, A., Lemoine, M., Moos, H.~W., 
        Sembach, K.~R., \& Jenkins, E.~B.\ 2002, \apjs, 140, 91 

	\bibitem[Woodgate et al.(1998)]{1998PASP..110.1183W} Woodgate, B.~E.~et 
	al.\ 1998, \pasp, 110, 1183 

      \bibitem[Woosley \& Weaver(1995)]{1995ApJS..101..181W} Woosley, S.~E.~\& 
	Weaver, T.~A.\ 1995, \apjs, 101, 181 

	\bibitem[Wright(1970)]{1970MNRAS.150..271W} Wright, M.~C.~H.\ 1970, \mnras, 
	150, 271 

        \bibitem[Wright \& Morton(1979)]{1979ApJ...227..483W} Wright, E.~L.~\& 
        Morton, D.~C.\ 1979, \apj, 227, 483 

\end{thebibliography}
\end{document}